\documentclass[twocolumn,tighten]{aastex61}

\usepackage{booktabs}

\received{July 1, 2016}
\revised{September 27, 2016}
\accepted{\today}
\submitjournal{ApJ}

\shorttitle{X-ray variability plane of AGN}
\shortauthors{O. Gonz\'alez-Mart\'in}

\begin{document}

\title{Update on the X-ray variability plane for active galactic nuclei: \\ The role of the obscuration}

\correspondingauthor{Omaira Gonz\'alez-Mart\'in, tenure track.}
\email{o.gonzalez@irya.unam.mx}

\author{Omaira Gonz\'alez-Mart\'in}
\affil{\centering Instituto de Radioastronom\'ia y Astrof\'isica (IRyA-UNAM), 3-72 (Xangari), 8701, Morelia, Mexico}




\begin{abstract}
Scaling relations are the most powerful astrophysical tools to set constraints to the physical mechanisms of astronomical sources and to infer properties that cannot be accessed directly. We re-investigate here one of these scaling relations in active galactic nuclei (AGN); the so-called X-ray variability plane \citep[or mass-luminosity-timescale relation,][]{McHardy06}. This relation links the power-spectral density (PSD) break frequency with the super-massive black hole (SMBH) mass and the bolometric luminosity. We used available \emph{XMM}-Newton observations of a sample of 22 AGN to study the PSD and spectra in short segments within each observation. This allows us to report for the first time that the PSD break frequency varies for each object, showing variations in 19 out of the 22 AGN analyzed. Our analysis of the variability plane confirms the relation between the break frequency and the SMBH mass and finds that the obscuration along the line of sight $\rm{N_H}$ (or the variations on the obscuration using its standard deviation, $\rm{\Delta N_H}$) is also a required parameter, at least for the range of frequencies analyzed here ($\rm{\sim 3\times10^{-5}-5\times 10^{-2}Hz}$). We constrain a new variability plane of the form: $\rm{log(\nu_{Break}) = (-0.589\pm 0.005) \,log(M_{BH}) + (0.10 \pm 0.01)\, log(N_{H}) - (1.5 \pm 0.3)}$ (or $\rm{log(\nu_{Break})= (-0.549\pm 0.009)\, log(M_{BH}) + (0.56 \pm 0.06)\, \Delta N_{H}+ (0.19 \pm 0.08)}$). The X-ray variability plane found by \citet{McHardy06} is roughly recovered when we use unobscured segments. We speculate that this behavior is well explained if most of the reported frequencies are related to inner clouds (within 1\,pc), following Kepler orbits under the gravitational field of the SMBH. 
\end{abstract}


\section{Introduction} \label{sec:intro}

Scaling relations describe strong trends that are observed between physical properties (such as mass, size, or luminosity) of galaxies. These relations are useful to set constraints to the physical mechanisms of astronomical sources and to infer properties for objects where they cannot be accessed directly \citep[e.g.][]{Faber76}. Various scaling relations have been found over the past decade between the black-hole (BH) mass ($\rm{M_{BH}}$) and the properties of the host galaxy; for instance the velocity dispersion \citep{Ferrarese00} or the mass \citep{Marconi03} of the galaxy bulges. The main implication is that we now believe that most galactic bulges harbor a super-massive BH (SMBH). The main application is that we can infer BH masses for systems with no direct measurements. 

Here we investigate one of these scaling relation associated to active galactic nuclei (AGN) variability, an ubiquitous property of AGN \citep[e.g.][]{Mushotzky93}. Several papers have demonstrated a strong, approximately linear relation between the timescale of the power-spectral density (PSD) break ($\rm{T_{Break} = 1/\nu_{Break} }$, where $\rm{\nu_{Break} }$ is the characteristic PSD break frequency) and the AGN BH mass, as expected from simple scaling arguments \citep[e.g.][]{Uttley02,Markowitz03,McHardy04,Vaughan05A}. \citet{McHardy06} found that $\rm{T_{Break}}$ depends on both BH mass and bolometric luminosity ($\rm{L_{bol}}$). This is the so-called X-ray variability plane of AGN. This scaling relation can also be used to infer BH masses \citep[e.g.][]{Porquet07,Gonzalez-Martin11}. Moreover, it could place a link between the physical process involved in AGN (with $\rm{M_{BH}\sim 10^{6}-10^{9} M_{\odot}}$) with BH X-ray binaries (BH-XRBs; $\rm{M_{BH}\sim 10M_{\odot}}$), with characteristic size scales and hence timescales simply scaling with their mass \citep{Shakura76}. 

However, the reliability of the X-ray variability plane is not clear yet \citep{Done05,Koerding07}. The most recent update on this relation was made by \citet{Gonzalez-Martin12} confirming the dependence of the break frequency on the BH mass. However, the bolometric luminosity was not statistically required in their analysis. Furthermore, the nature of this relation is not well understood. It was firstly suggested that the break frequency might correspond to the viscous or thermal time-scales at a few gravitational radius \citep[i.e. related to the accretion disk radius,][]{McHardy04}. However, \citet{Zhang17} recently proposed that the break frequency could be the result of obscuration variations due to clouds in the broad line region (BLR). 

When it comes to timescales, the BH mass and the bolometric luminosity are extremely different AGN parameters. The BH mass can be considered as a constant value per object because we do not expect significant mass changes within timescales of years. On the contrary, most AGN are variable at X-rays in long (months to years) and short (minutes or hours) time scales \citep{Markowitz03,Gonzalez-Martin12}. Therefore, if the timescale of the PSD break depends on the luminosity, it should vary on short timescales. Thus, the hypothesis (reinforced by our results) is that the break frequency should vary for a single object. An average bolometric luminosity and break frequency over a few tens up to hundreds of ksec (as traced by a single \emph{XMM}-Newton) could smear out any dependence of the X-ray variability plane. 

This paper study the X-ray PSDs and spectra of a sample of 22 bright AGN in short periods of time (down to 10 ksec) to better constrain the AGN variability plane. The paper is organized as follows: Section \ref{sec:sample} shows the sample selection, Section \ref{sec:data} includes the data processing and analysis, Section \ref{sec:results} presents the main results of the analysis, Section \ref{sec:discussion} discuss the use and implication of the new X-ray variability plane and, finally Section \ref{sec:summary} includes a summary of the main results. Throughout this paper, we assumed the Hubble constant as $\rm{H_{0}=70 km/s/Mpc}$.

\begin{table*}[tbh]
\scriptsize
\begin{center}
\caption{Main properties of the AGN sample}
$
\begin{array}{lllccccccccc}
\toprule
Objname &Type & log(M_{BH}) & N_{obs} & Expos. & count$-$rate & bin & log(L_{bol}) & log(N_{H}) & log(\nu_{Break})   & N_{Break}/N_{N_{H}}      \\
			&               &                &                 &   (ksec)  &   (counts/s) &   (s) &        &       &                &            \\
\midrule
MRK\,335 				& NLSy1		& 7.23\pm0.04^{1} (R) 	& 	4 	& 389	& 0.96 &	50 	& [43.7, 44.8]  &  [20.8,21.9]   & [-3.9,-2.6] &  116/8  \\
ESO\,113$-$G010 			& Sy1		& 						& 	1	& 93	& 0.34 &100 & [43.4,43.6]  &		21.5		& [-3.7,-2.9] &   42/1    \\
Fairall\,9 					& Sy1		& 8.3\pm0.1^{1} (R)		& 	5	& 332	& 1.45 &	50 	& [44.7,45.1]  &		21.1		&     -3.6	      &   1/0  	 \\
PKS\,0558$-$504 		& NLSy1		& 8.48\pm0.05^{2} (RP)		& 	5	& 564	& 1.11 &	50 	& [45.7,46.4]  &	  [20.9,21.4]	& [-4.0,-2.8] &  69/0     \\
1H\,0707$-$495 			& NLSy1		& 6.3\pm0.5^{3}	(L)		& 	14	&1095 	& 0.12 &200 & [42.7,43.8]  &  [21.1,22.2]	 & [-3.5,-2.7] &   58/4   \\
ESO\,434$-$G40 			& Sy1		& 7.57\pm0.25^{4} (S)	&	 5	& 382	& 7.62 &	50 	& [44.0,44.3]  &  [21.8,22.2]	 & [-4.1,-2.9] &  232/232      \\
NGC\,3227 					& Sy1		& 6.8\pm0.1^{1} (R)		& 	2	& 124	& 2.02 &	50 	& [42.4,43.0]  &  [20.7,22.6]	 & [-3.6,-2.5] &   46/43     \\
REJ\,1034$+$396 		& NLSy1		& 6.6\pm0.3^{5}	(L)		& 	7	& 277	& 0.12 &	200 & [43.3,43.8]  &	[21.4,21.9]     & [-3.5,-2.9] &   4/0      \\
NGC\,3516 					& Sy1		& 7.40\pm0.05^{1} (R) 	& 	6	& 440	& 2.77 &	50 	& [43.6,44.3]  & [20.6,21.9]	 & [-4.5,-2.8]  &   47/37     \\
NGC\,3783 					& Sy1		& 7.37\pm0.08^{1} (R)	& 	3	& 223	& 3.81 &	50 	& [43.9,44.2]  & [20.6,21.5]	& [-4.4,-3.4] &   28/5       \\
NGC\,4051 				& NLSy1		& 6.1\pm0.1^{1}	(R)		& 	13	& 435 	&1.12 & 50 	& [41.8,42.7]  & [20.9,22.1] 	& [-3.6,-2.2] &   63/12    \\
NGC\,4151 					& Sy1		& 7.55\pm0.05^{1} (R)	& 	13	& 440	& 3.73 &	50 	& [42.7,43.5]	  & [22.2,23.1]	& [-4.0,-2.9 &   9/9       \\
MRK\,766 				& NLSy1		& 6.2\pm0.3^{6}	(R)		& 	9	& 596	& 1.21 &	50 	& [43.1,43.9]	  & [20.9,21.9]	& [-3.7,-2.4] &  257/44     \\
NGC\,4395					& Sy1		& 5.4\pm0.1^{7}	(R)		& 	3	& 175	& 0.38 &	100 & [40.9,41.1]	  & [21.0,22.8]	& [-3.2,-2.4] &   70/53     \\
MCG$-$06$-$30$-$15	& NLSy1		& 6.3\pm0.4^{5}	(L)		& 	7	& 563	& 3.06 &	50 	& [43.2,43.9]	  & [20.5,21.7]	& [-3.7,-2.6] &  336/81     \\
IC\,4329A 					& Sy1		& 8.3\pm0.5^{8}	(S)		& 	1	& 125	& 7.41 & 50	& [44.6,44.7]	  & [20.5,21.4]	  & [-4.4,-3.0] &   51/48      \\
Circinus 					& Sy2		& 6.04\pm0.08^{9} (M)	& 	4	& 190	& 0.57 &	50 	& [41.3,41.4]	  &	[21.4,22.0]	 & [-4.2,-4.2] &   11/0    \\
NGC\,5506 				& NLSy1		& 8.1\pm0.2^{10}	 (R)		& 	3	& 276	& 5.87 &	50 	& [43.4,43.9]	  & [22.4,22.6] 	 & [-3.9,-2.8] &  110/110  \\
NGC\,5548 					& Sy1		& 7.72\pm0.02^{1} (R)	& 	8	& 349	& 2.10 &	50 	& [44.3,44.8]	  & [20.8,22.3] 	 & [-4.2,-2.9] &   17/4     \\
NGC\,6860 					& Sy1		& 7.6\pm0.5^{11}	 (L)		& 	1	& 88	& 2.12 &	50	 & [43.8,44.0] &[20.8,22.0] 	 & [-4.0,-3.7] &   48/48 \\
ARK\,564 				& NLSy1		& 6.3\pm0.5^{12}	 (S)		& 	9	& 494	& 1.78 &	50 	& [44.0,44.7]	  &	        21.1              & [-3.1,-2.2] &  398/1    \\
NGC\,7469 					& Sy1		& 6.96\pm0.05^{1} (R)	& 	9	& 719	& 2.31 &	50 	& [43.8,44.1]	  &	 [20.7,21.3]      & [-4.1,-2.7] &   50/0     \\
\bottomrule
\end{array}
$
\end{center} 
\tablecomments{Type: AGN classification (NLSy1 -- Narrow-line Seyfert 1; Sy1 -- Seyfert 1; Sy2 -- Seyfert 2); $\rm{N_{obs}}$: number of exposures; Expos.: total net exposure time per object; $\rm{count-rate}$: average count-rate per source; $\rm{bin:}$ selected bin to extract light-curves (depending on the average count-rate, see text); $\rm{log(L_{bol})}$: minimum and maximum bolometric luminosity (in erg/s) obtained from the X-ray luminosity (see text) in logarithmic scale; $\rm{log(N_H)}$: minimum and maximum hydrogen column density in logarithm scale; $\rm{log(\nu_{Break})}$: minimum and maximum PSD break frequency in logarithmic scale; $\rm{N_{Break}}$: number of detected PSD break frequencies; and $\rm{N_{N_{H}}}$: number of detected absorptions among the detected PSD break frequencies.  BH mass estimates from: (1) \citet{Zu11}; (2) \citet{Gliozzi10}; (3) \citet{Bian03}; (4) \citet{Peng06}; (5) \citet{Zhou10}; (6) \citet{Bentz09}; (7) \citet{Peterson05}; (8) \citet{Markowitz09}; (9) \citet{Graham08}; (10) \citet{Du15}; (11) \citet{Wang07}; (12) \citet{Zhang06}.}
\label{tab:sample}
\end{table*}

\section{Sample} \label{sec:sample}

\begin{figure*}
\begin{center}
\includegraphics[width=2.0\columnwidth]{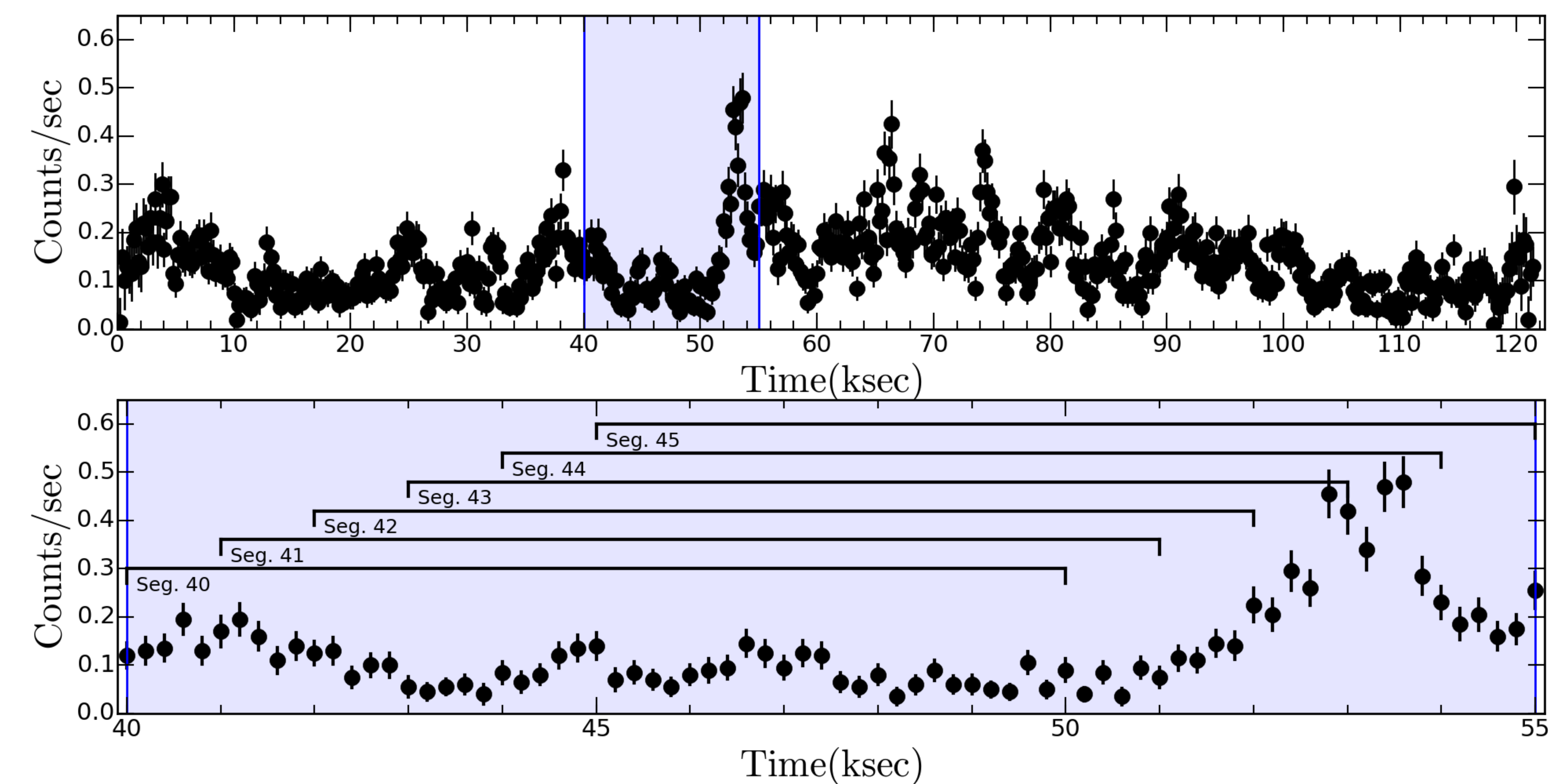}
\caption{(Top panel): Light curve of 1H\,0707-495 (ObsID 0511580101) in the 2-10 keV band with 200 s time bins. (Bottom panel): Zoom to the light curve in the bottom panel in the range between 40 to 55 ksec (also displayed as a shaded area in the top panel). Horizontal brackets show six segments used to obtain PSDs and spectra to illustrate the methodology applied to the data.}
\label{fig:1Hlightcurves}
\end{center}
\end{figure*}

Our sample includes all the AGN with reported PSD break frequencies in \citet{Gonzalez-Martin12}. This is the largest compilation of PSD break frequencies, including 15 detections with \emph{XMM}-Newton data by \citet{Gonzalez-Martin12} and seven previously detected with RXTE monitoring data \citep[see Table~4 in][]{Gonzalez-Martin12}. This sample is the best suited for our analysis because it guarantees that the PSD break frequency has been measured before for them. 

We collected all the \emph{XMM}-Newton data available in the Heasarc archive\footnote{https://heasarc.gsfc.nasa.gov} with exposure times longer than 20 ksec. We imposed this exposure time limit to be able to produce several segments of $\rm{\sim}$10 ksec. Table~\ref{tab:sample} includes the main properties of the sample. The observational details reported in this table are the AGN type, BH mass, number of exposures analyzed, total exposure time, and average count-rate (Col.~2-6). BH masses are taken from reverberation mapping techniques (R), m-sigma relation (S), continuum at 5100$\rm{\AA}$ (L), water masers (M), and radio fundamental plane (RP). No accurate BH mass estimate has been reported for ESO\,113-G010 (discussed in Section \ref{sec:discussion}). Several BH masses are reported for individual objects (see Table \ref{tab:BHmass} for a compilation of available BH masses). Appendix \ref{sec:BHmass} gives details on the selection criteria for the BH masses used throughout this paper.

This sample includes 22 AGN (1 Seyfert 2, 12 Seyfert 1, and 9 narrow-line Seyfert 1). The sample expands six orders of magnitudes in bolometric luminosities ($\rm{L_{bol} \simeq 7.7\times 10^{40}-2.5 \times 10^{46}}$ erg/s, Col.~8 in Table~\ref{tab:sample}) and more than three orders in BH masses ($\rm{M_{BH}\simeq 2\times 10^5-3\times 10^{8} M_{\odot}}$). We have collected almost 100 days of \emph{XMM}-Newton observations for the entire sample. 

\section{Data processing and analysis}\label{sec:data}

We use data from the EPIC pn camera \citep{Struder01} of \emph{XMM}-Newton. We did not include data from MOS camera to avoid cross-calibration issues that might introduce unknown uncertainties on the timing and spectral analysis. These objects are quite bright in X-rays so the EPIC pn camera alone is enough to have high sensitivity spectra and light-curves. The data were reduced with SAS v15.0.0, using the most up-dated calibration files available (at April 2017). 

Nuclear positions were retrieved from NED. We used circular regions with 25 arcsec radii (500 pixels) to extract spectra and light-curves of the targets. This circular region encircles 80\% (85\%) of the PSF at 1.5 keV (9.0 keV) for an on-axis source with the EPIC pn instrument. The background events were selected from a source-free circular region on the same CCD as the source. We selected only single and double pixel events (i.e. PATTERN==0-4). Bad pixels and events too close to the edges of the CCD chip were rejected (using the standard FLAG==0 inclusion criterion).

Background flares could dominate the observation. We built an event list of high background to take them into account in our analysis. For this purpose, we extracted a light-curve of the background of the observations (i.e. masking any source in the event file) with energies above 10 keV. We selected those periods with a background light-curve exceeding three times the standard deviation over the mean. We recorded them in an event file to debug our final light-curves. Hereinafter we refer to this file of high count rate in the background as background flare file (BGF).

All the observations were made with the small window and fast readout mode to avoid pile-up effects, which is below 10\% in our sample. Thus, no special treatment for pile-up was considered.  

\subsection{Timing analysis}\label{sec:timing}

The source and background light curves were extracted using {\sc evselect} at the 2-10 keV energy band. We used 50 s time bins except for four low count-rate sources where we chose a bin size of 100 or 200 s (see Col.~7 in Table~\ref{tab:sample}). Any interval with high background periods recorded in the BGF file (see above) were excluded from this analysis. 

We produced segments of the light-curves to detect PSD variations in short time intervals. We produced light curves of 10, 20, 40, 60, 80, and 100 ksec every 1ksec if the length of the observation allowed it. Fig.~\ref{fig:1Hlightcurves} shows one of the light curves of 1H\,0707-495, giving a sketch of the 10 ksec segments made. The first segment shown includes the temporal range between 40 up to 50 ksec while the second segment includes the range between 41 up to 51 ksec. Thus, each segment includes a (large) fraction of the segments next to it. We have found that the PSD break frequency changes fast within a single \emph{XMM}-Newton observation (see below). Thus, this methodology is designed to get the highest number of break frequencies with accurate measurements of the spectral properties. We chose several segments lengths (from 10 up to 100 ksec) to detect relatively low break frequencies. For a particular time, we gave priority to the break frequencies detected with the shortest segment to ensure the best match of the break frequency with the spectral properties. This is crucial to characterize the X-ray variability plane. This procedure is sensitive to PSD break frequencies in the range $\rm{\nu_{Break}\sim [10^{-5},10^{-2}]}$~Hz. 

\begin{table*}[t!]
\scriptsize
\begin{center}
\caption{Variability plane results}
$
\begin{array}{lllcccccccc}
\toprule
Model 					&	N_{seg}		&Av.	&		A                        & B                        & C                    & D /D^{\star}                & E   & r  & \chi^{2} /dof  & ftest\\
							& 			&			&		(log(M_{BH}))    &  (log(L_{bol}))    &    (\Gamma)  &(log(N_{H})/\Delta(N_{H}))  &       &   &   &  \\
\midrule \midrule
Alog(M_{BH}) + E									& 2021 & o &  $-$0.409 \pm 0.005   &         \dots                 &       \dots        &        \dots     &   $-$0.65 \pm 0.04  & 0.67   & 7434.1/18 &   \\
               										&         & \nu &   $-$0.81  \pm 0.01   &           \dots         &         \dots       &       \dots      & 2.26 \pm 0.09   & 0.87  &  250.8/14     &  \\
               										& 2009 & o &  $-$0.408 \pm 0.005  &         \dots         &          \dots       &         \dots     &  $-$0.65 \pm 0.04  & 0.67  & 7438.5/18  &   \\
               										&         & \nu &   $-$0.81 \pm 0.01    &          \dots         &         \dots       &     \dots      & 2.26 \pm 0.09   & 0.88  & 293.3/14    &  \\
               										&  739  & o &   $-$0.396 \pm 0.006 &           \dots       &          \dots   &      \dots     &  $-$0.59 \pm 0.04  & 0.78  &   265.8/13 &  \\
               										&         & \nu &  $-$0.595 \pm 0.005  &           \dots      &        \dots      &      \dots     & 0.75 \pm 0.04   & 0.97  &   78.7/13   &  \\ \midrule
Alog(M_{BH}) + B log(L_{bol}) + E						& 2009 & o &     $-$0.610 \pm 0.008     &    0.193 \pm 0.005   &       \dots       &      \dots    &    $-$7.6 \pm 0.2 &  0.73 &  4311.3/17  &  0.003  \\
               										&         & \nu &   $-$0.84   \pm 0.02    &      0.07  \pm 0.03    &           \dots     &        \dots    &  $-$0.9 \pm 1.5  & 0.88  &  11.9/13    & \checkmark \\
               										&  739  &o  &    $-$0.41  \pm 0.01    &     0.019  \pm 0.009    &       \dots      &     \dots     & $-$1.3  \pm 0.3   & 0.78  &  231.5/12    &  0.2 \\
               										&         & \nu &    $-$0.63  \pm 0.02  &     0.03  \pm 0.02    &       \dots      &       \dots    &  $-$0.3  \pm 0.5  & 0.97  &   39.8/12  &  0.005 \\ \midrule
Alog(M_{BH}) + C\Gamma + E						& 2009 & o &   $-$0.430 \pm 0.005     &     \dots         &     0.297 \pm 0.009    &      \dots     &  $-$0.99 \pm 0.04  & 0.74  &   3422.4/17  &  0.0003 \\
               										&         & \nu &   $-$0.48  \pm 0.04    &         \dots       &    1.3   \pm 0.1   &       \dots      &  $-$2.5  \pm 0.5  & 0.92  &   25.2/13  & \checkmark  \\
               										&  739  & o  &   $-$0.401 \pm 0.005  &         \dots          & $-$0.11 \pm 0.01    &       \dots    & $-$0.37 \pm 0.04   &  0.79 &   144.0/12 &  0.008  \\
               										&          &\nu &   $-$0.60  \pm 0.01  &          \dots        &  0.03 \pm 0.04   &       \dots     &  0.74 \pm 0.04  &  0.97 &   27.2/12  & 0.0005   \\ \midrule
Alog(M_{BH}) + D log(N_{H}) + E						&  739  & o &  $-$0.410 \pm 0.006   &       \dots           &       \dots      &   0.13 \pm 0.01   & $-$3.3 \pm 0.2   & 0.89  &  56.5/12  & \checkmark   \\
               										&         & \nu &  $-$0.589 \pm 0.005   &        \dots          &      \dots      &    0.10 \pm 0.01  &  $-$1.5 \pm 0.3  & 0.97  &  6.6/12  &   \checkmark  \\  \midrule
Alog(M_{BH}) + D^{\star} \Delta(N_{H}) + E			&  739  & o &  $-$0.30 \pm 0.01   &       \dots           &       \dots      &   0.9 \pm 0.1   & $-$1.4 \pm 0.1   & 0.89  &  66.0/12  & \checkmark   \\
               										&         & \nu &  $-$0.549 \pm 0.009   &        \dots          &      \dots      &    0.56 \pm 0.06  &  0.19 \pm 0.08  & 0.98  &  3.7/12  &   \checkmark  \\ 
\midrule \midrule
\multicolumn{10}{c}{Segments~without~absorption} \\ \midrule
Alog(M_{BH}) +  E								& 1270 &\nu & $-$1.03 \pm 0.03 &	  \dots 	&  \dots 	& 	  \dots 	&	3.5  \pm 0.2	&	0.84	&	851.2/14 	& 		\\
Alog(M_{BH}) + B log(L_{bol}) + E					& 1270 &\nu &$-$1.39  \pm 0.05 &	0.82  \pm 0.08&	  \dots 	&   \dots 	&$-$30.0  \pm 3.5&	0.92	&	69.5/13 	& 	\checkmark	\\
Alog(M_{BH}) + C\Gamma + E						& 1270 &\nu &$-$0.70 \pm 0.03 &	  \dots 	&1.22 \pm 0.05& 	  \dots 	&$-$1.07 \pm 0.27	&	0.95	& 79.3/13 & \checkmark		\\
\midrule \midrule
\multicolumn{10}{c}{Objects~without~absorption~variations} \\ \midrule
Alog(M_{BH}) +  E								& 1423 &\nu &$-$0.270 \pm 0.006	&	 \dots		 &	 \dots			& 	 \dots	&	$-$1.47 \pm 0.04 &	0.46	 & 15696.3/13	& 		\\
Alog(M_{BH}) + B log(L_{bol}) + E					& 1423 &\nu &$-$0.84 \pm 0.02	&	0.57  \pm 0.01 &	 \dots			& 	 \dots	&	$-$22.7 \pm 0.4	 &	0.89	 & 329.6/12	& \checkmark \\
Alog(M_{BH}) + C\Gamma + E						& 1423 &\nu &$-$0.287 \pm 0.006	&	 \dots		 &	0.894 \pm 0.009	& 	 \dots	&	$-$3.18 \pm 0.04 &	0.86	& 582.2/12	& \checkmark \\

\bottomrule
\end{array}
$
\end{center} 
\tablecomments{ Col.~1 shows the baseline model used; Col.~2 gives the number of segments involved for each fit (it depends on the availability of $\rm{M_{BH}}$, $\rm{L_{bol}}$, $\rm{\Gamma}$, $\rm{N_H}$ values for the detected PSD frequency breaks, see text); Col.~3 shows if the results are obtained by averaging per object (denoted as `o') or by intervals of break frequencies (denoted as `$\rm{\nu}$'); Cols.~4-8 give the parameters of the fit; Cols. 9 and 10 shows the resulting correlation coefficient and the $\rm{\chi^{2}}$ statistic over the degree of freedom (dof), respectively; and Col.~11 shows the resulting f-test statistic. Note that `\checkmark' is shown when f-test probability is below $\rm{10^{-4}}$. Col. 7 shows the results for both D and $\rm{D^{\star}}$ corresponding to Eqs. \ref{eq:bhnh} and \ref{eq:bhnhvar}, respectively. We repeated the analysis (below double lines) using only segments without absorption and objects without recording absorption variations to try to recover any dependency of the break frequency on the bolometric luminosity or the spectral index (see text). }
\label{tab:plane}
\end{table*}

The PSD gives the distribution of variability power (amplitude squared) as a function of the temporal frequency. PSD can be obtained by calculating the periodogram \citep{Vaughan03}. The periodogram data were fitted using the maximum likelihood method discussed in \citet{Vaughan10}. \citet{Gonzalez-Martin12} fitted the PSDs to two models. The first one is a single power-law model:
\begin{equation}
P(\nu)=N\nu^{-\alpha} + C
\end{equation}

\noindent where the free parameters are the normalization N, the slope $\rm{\alpha}$, and the constant C (associated to the Poisson noise). The second model is a bending power-law model:
\begin{equation}
P(\nu)=N\nu^{-\alpha_1}\Big(1+\Big\{\nu / \nu_{Break} \Big\}^{\alpha_2-\alpha_1}\Big)^{-1} + C 
\end{equation}

\noindent where the free parameters are the normalization N, the slope above the break frequency $\rm{\alpha_2}$, the break frequency $\rm{\nu_{Break}}$, and the constant C. Following \citet{Gonzalez-Martin12}, we fixed the slope before the break frequency $\rm{\alpha_1=1}$ because X-ray long-term monitoring of AGN show that $\rm{\alpha_1=1}$ is appropriate in several AGN \citep[][]{Vaughan05A}. However, other slopes before the break are expected by analogy with BH-XRBs \citep[e.g.][]{McHardy03}. Although we do not report it here, we repeated the analysis with a set of slopes before the break frequency. Interestingly, almost flat slopes might be relevant for a sizable amount of the segments in our sample which are not temporary coincident with the break frequencies found for $\rm{\alpha_1=1}$. For some of them, low and high frequency breaks have been found simultaneously. Indeed, among the objects in our sample ARK\,564 is known to show this behavior \citep{Papadakis07,McHardy07}. We are working in a second publication to investigate these results (Gonzalez-Martin in prep.). However, since this is out of the scope of this paper, we chose for this study all the frequency breaks detected with $\rm{\alpha_1=1}$ because it provides the maximum number of detected frequency breaks. 

\begin{figure*}
\begin{center}
\includegraphics[width=0.69\columnwidth]{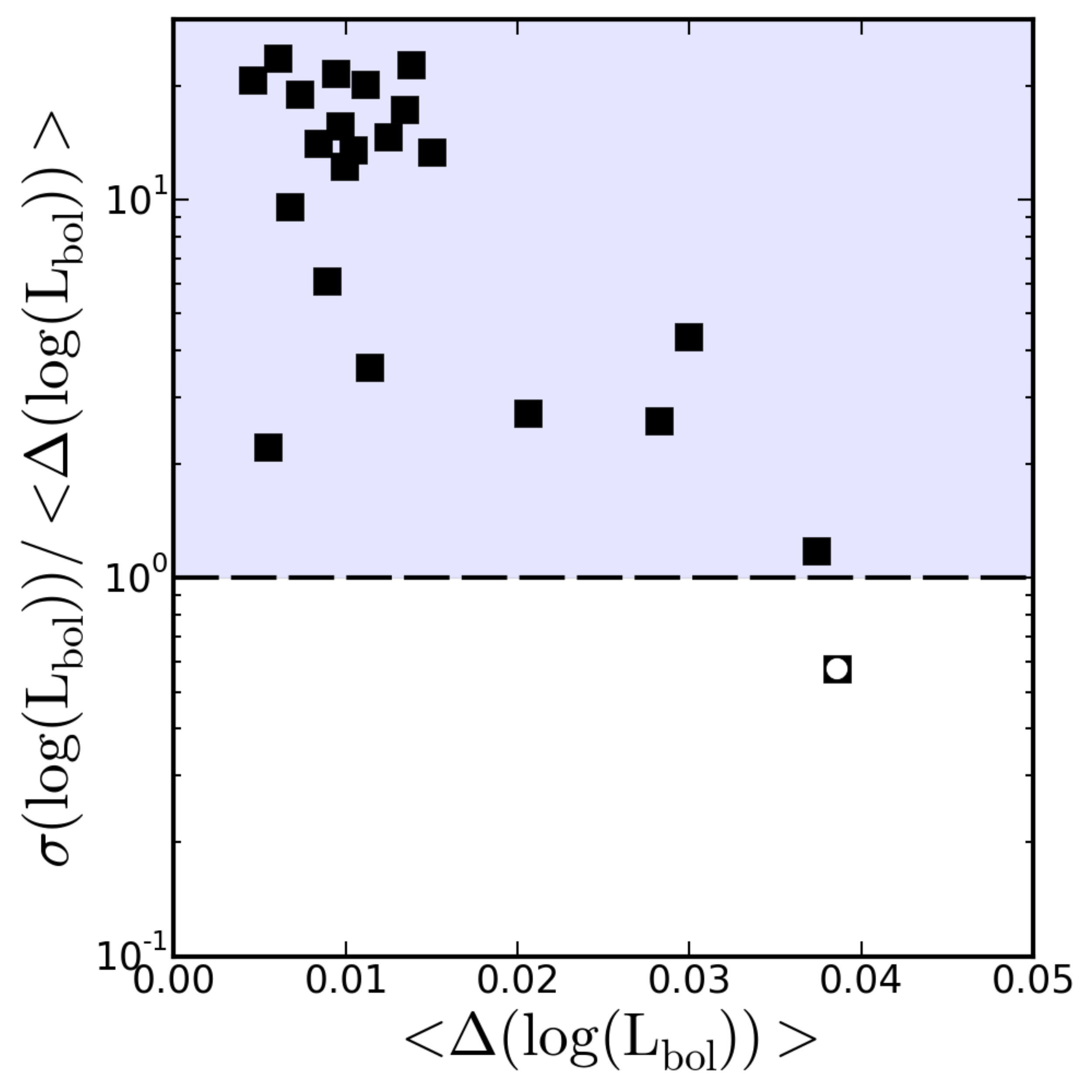}
\includegraphics[width=0.69\columnwidth]{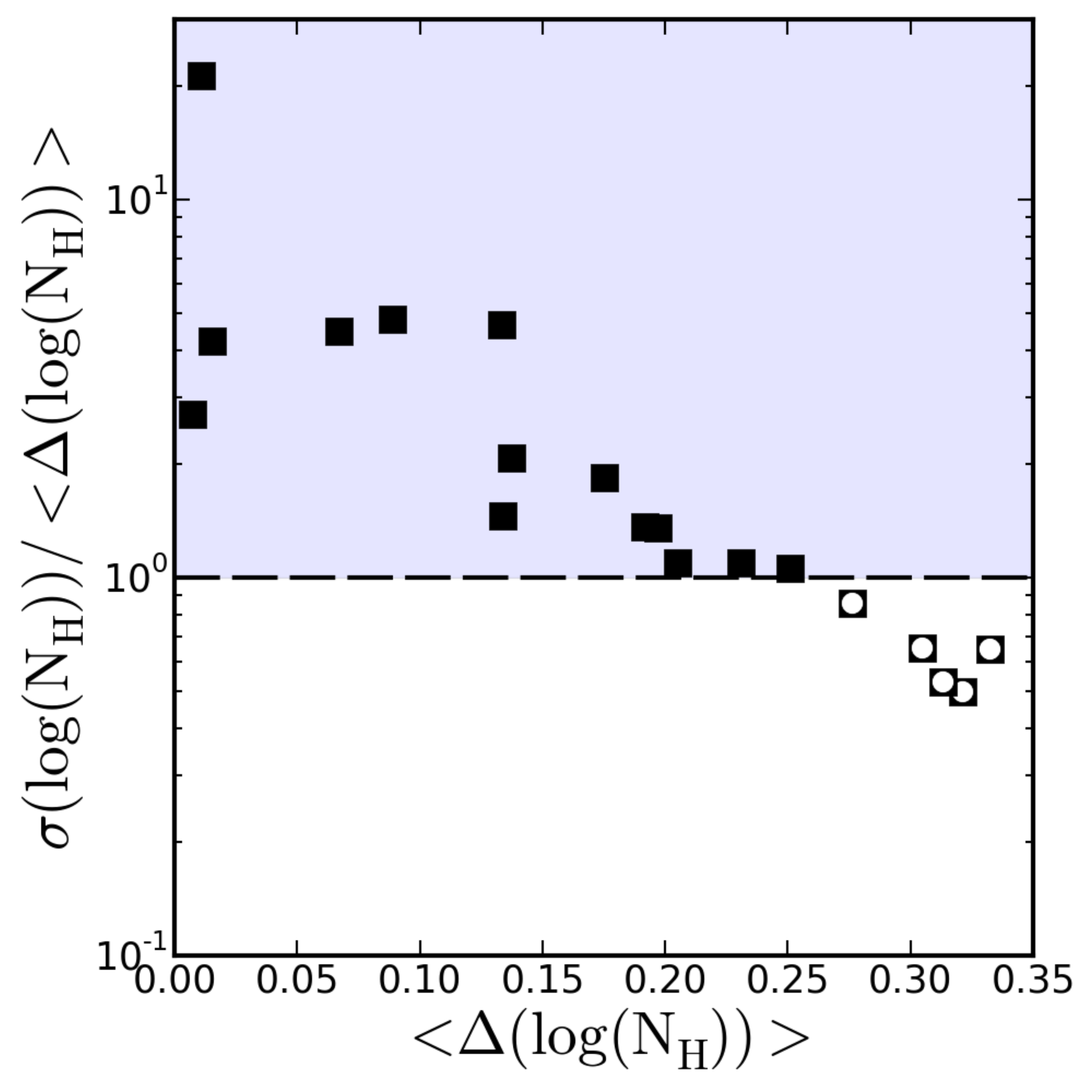}
\includegraphics[width=0.69\columnwidth]{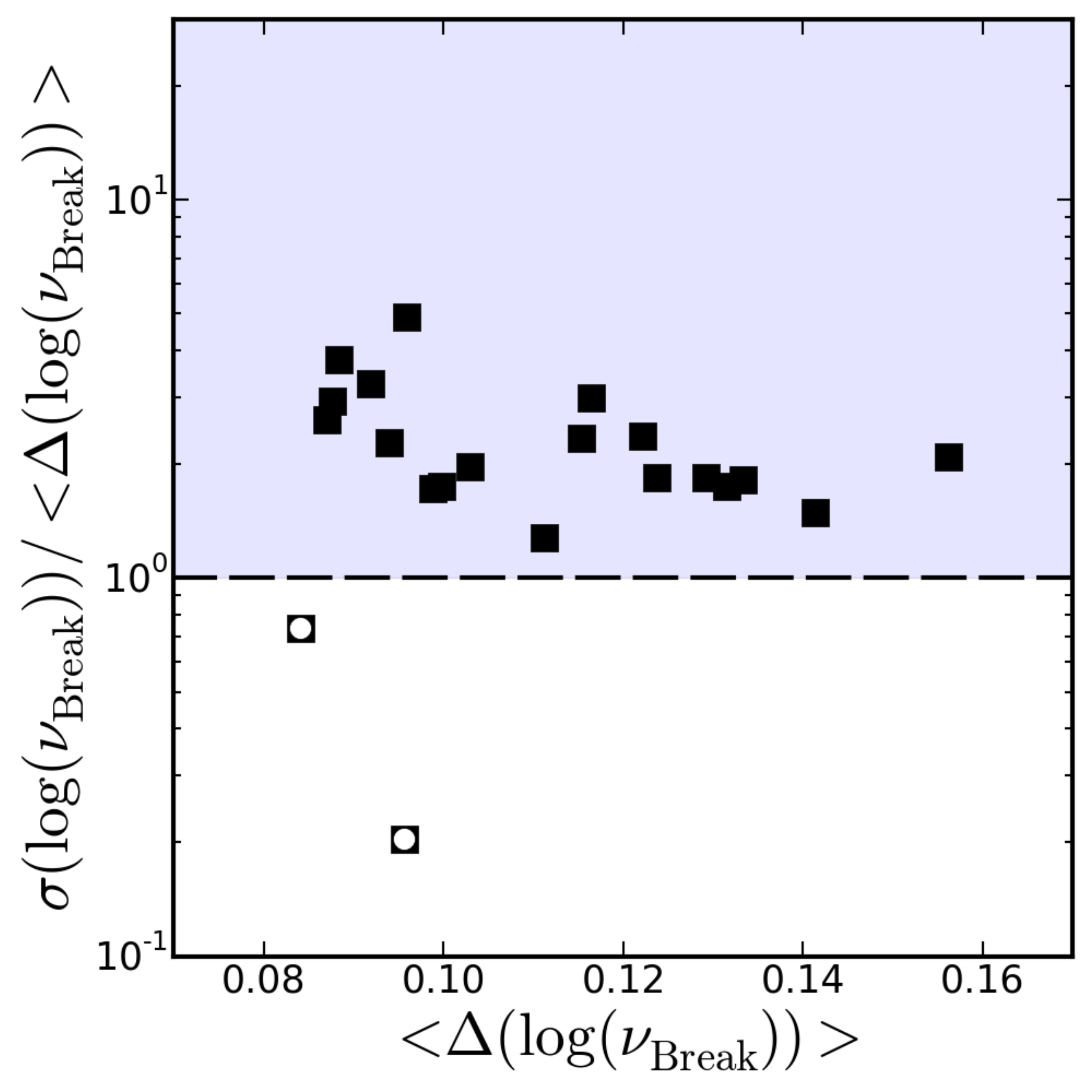}
\caption{Standard deviation over the average error versus its average error for the bolometric luminosity (left panel), hydrogen column density (middle panel), and PSD break frequency (right panel). Objects above the dashed line are variable. White circles highlight not variable objects. Note that ESO\,113-G010, Fairall\,9, and ARK\,564 are excluded from the middle panel due to the lack of enough $\rm{N_{H}}$ measurements. Furthermore, Fairall\,9 is neither included in the right panel due to the lack of enough detected frequency breaks.}
\label{fig:variability}
\end{center}
\end{figure*}

We used the likelihood ratio test to determine when the bending, power-law model is preferred against the single, power-law model with a significance threshold of $\rm{p <0.01}$. Unfortunately, in most of the cases the slope after the frequency break (i.e. $\rm{\alpha_{2}}$) is poorly constrained even when the bending power-law model is preferred. This slope is not used in the subsequent analysis. 

We found 2063 break frequencies in individual intervals of time in the set of AGN presented in this analysis. The break frequency range and the number of break frequencies detected per object are included in Col.~10 and 11 of Table~\ref{tab:sample}. Col.~11 also reports the number of segments where both break frequency and absorption are measured. 

\subsection{Spectral analysis}\label{sec:spectral}

Individual spectra were extracted using the same temporal segments used for the timing analysis. Regions were extracted by using the {\sc evselect} task and redistribution matrix, and effective areas were calculated with {\sc rmfgen} and {\sc arfgen} tasks, respectively. We also excluded in the subsequent analysis any segments with high background periods included in the BGF file. 

Bright Seyfert 1 and NLSy1 galaxies analyzed in this paper correspond to the thermal soft states in BH-XRBs, where an optically thick and geometrically thin disc is assumed to extend down to the innermost stable circular orbit around a SMBH \citep{Gierlinski08}. The hard X-ray photons in AGN are originated in a corona closely linked to the accretion disc. Under this scenario, we fitted the 2--6 keV spectra to an absorbed power-law model. We excluded energies below 2 keV to avoid other components not related to the intrinsic continuum of the AGN and energies above 6 keV to avoid the complexity of the FeK$\rm{\alpha}$ line and large contributions of the reflection component. This model includes an absorber accounting for the intrinsic obscuration of the source. The free parameters of this model are: the hydrogen column density $\rm{N_{H}}$, the spectral index $\rm{\Gamma}$, and the normalization of the power-law. The spectral analysis was performed using XSPEC (version 12.9.0i). Power-law model has a well known degeneracy between the spectral index and the hydrogen column density \citep{Suchy08}. We compute two dimensional confidence contours to confirm this degeneracy is not responsible for the variations seen for these parameters. Considering the importance on the hydrogen column density estimates in this paper (see Section \ref{sec:results}), we produced 10,000 simulated spectra to trace the sensitive range of $\rm{N_{H}}$ with our spectral fitting. The result is that this method is sensitive to the range between $\rm{4\times10^{20}- 4\times 10^{23} cm^{-2}}$, recovering more than 99\% of the $\rm{N_{H}}$ values with less than 5\% error. This range covers the expected range of absorptions associated to BLR \citep{Netzer90, Peterson97}. More fundamental discrepancies on the parameters could be found if our model is not the underlaying model for these sources. This is discussed in Section \ref{sec:discussion_nh}.

We then computed the intrinsic (i.e. absorption corrected) 2-10 keV luminosity. We used this intrinsic X-ray luminosity to infer the bolometric luminosity of the source using the relation given by \citet{Marconi04}: 
\begin{equation}
\footnotesize
log(\mathcal{L}/L(2-10~keV))=1.54+0.24\mathcal{L}+0.012\mathcal{L}^{2}-0.0015\mathcal{L}^3
\end{equation}

\noindent where $\rm{\mathcal{L}}\rm{=(logL_{bol}-12)}$ and $\rm{L_{bol}}$ is in units of $\rm{L_{\odot}}$. The range of bolometric luminosities and hydrogen column densities per object is given in Col.~8 and 9 in Table~\ref{tab:sample}.

\section{Results}\label{sec:results}

\subsection{Variability per object}\label{sec:perobject}

We investigate in this section if the sources in our sample show significative variations on the bolometric luminosity $\rm{L_{bol}}$, hydrogen column density $\rm{N_H}$, and PSD break frequency $\rm{\nu_{Break}}$. To do so, we compare the standard deviation over the mean versus its average error in Fig.~\ref{fig:variability}. Most of the objects in our sample show variations in the three parameters (14 out of the 22 objects). Circinus galaxy shows no evidence of variations in none of the parameters. All but three objects (namely Fairall\,9, Circinus, and NGC\,6860) show variations in the frequency break. Eight sources have detected breaks without any signs of $\rm{N_{H}}$ variations, namely ESO\,113-G010, Fairall\,9, PKS\,0558-504, 1H\,0707-495, REJ\,1034+396, Circinus, ARK\,564, and NGC\,7469. 

We do not see a bias toward less variable sources among those with larger error bars in the break frequency while objects with larger errors on the bolometric luminosity and hydrogen column density are less variable (Fig.~\ref{fig:variability}). Thus, the lack of variability found in these two parameters could be due to the lack of good constraints on the parameters rather than no variation at all. 

\subsection{X-ray variability plane}

We use the set of detected break frequencies in our sample to perform the analysis of the X-ray variability plane. For that purpose, we investigated five plausible variability planes. The first one is the well known relation between the PSD break frequency and the BH mass (hereinafter \emph{break-mass} relation): 
\begin{equation}
log(\nu_{Break})= A~log(M_{BH}) + E \label{eq:bh}
\end{equation}

\noindent This plane is well established \citep[][]{McHardy06,Gonzalez-Martin12} so we use it to test the statistical significance of more complex relations. \citet{McHardy03} proposed that the relationship between break frequency and BH mass is not the same for all AGN. They propose that there is (at least) another underlying parameter which governs the location of the break frequency/mass relation so that it may be different for galaxies of different types. They showed that for a given BH mass, the break frequency is smaller in NLSy1s. Based on this, \citet{McHardy06} proposed a relation between the break frequency, the BH mass, and the bolometric luminosity (hereinafter \emph{break-mass-luminosity} relation): 
\begin{equation}
log(\nu_{Break})= A~log(M_{BH}) +B~log(L_{bol}) + E \label{eq:bhlbol}
\end{equation}

\noindent The spectral index should be tightly connected to the bolometric luminosity because there is a well known relation between these two parameters \citep{Shemmer06}. We included the spectral index to test if it could better recover the underlying relation to the break frequency. Thus, the third plane replaces the bolometric luminosity in the former equation by the spectral index (hereinafter \emph{break-mass-index} relation): 
\begin{equation}
log(\nu_{Break})= A~log(M_{BH}) +C~\Gamma + E \label{eq:bhgamma}
\end{equation}

\citet{Zhang17} recently suggested that the break frequency could be related to eclipses, along the line of sight to the observer rather than the accretion process. Under this scenario, we might expect a relation between the break frequencies and the absorption. Thus, the last plane tested relates the break frequency to the absorption along the line of sight to the observer (hereinafter \emph{break-mass-nh} relation):
\begin{equation}
log(\nu_{Break})= A~log(M_{BH}) +D~log(N_{H}) + E \label{eq:bhnh}
\end{equation}

\noindent Note that this last variability plane has never been tested before. A slight variation of this latter plane contemplates variations on the absorption along the line of sight rather than the absorption itself (hereinafter \emph{break-mass-nhvar} relation):
\begin{equation}
log(\nu_{Break})= A~log(M_{BH}) +D^{\star}~\Delta(N_{H}) + E \label{eq:bhnhvar}
\end{equation}

\noindent where $\rm{\Delta(N_{H})}$ is the standard deviation (for a single object or per break frequency bin, see below) of the absorption. We denote the constant associated with the absorption and absorption variations in the last two relations as \emph{D} and $D^{\star}$, respectively, to stress that these two planes are both associated with absorption. 

\begin{figure*}
\begin{center}
\includegraphics[width=0.67\columnwidth]{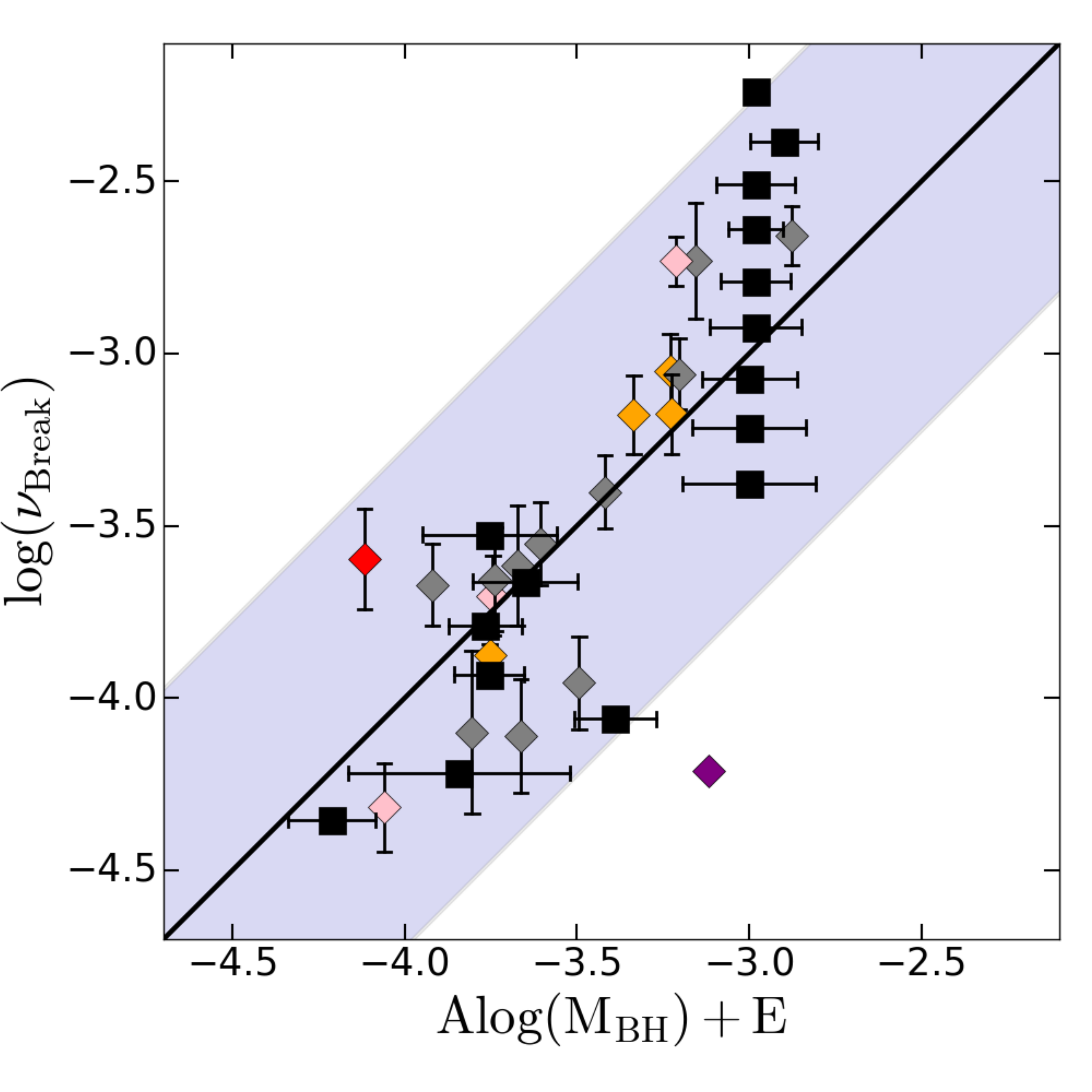}
\includegraphics[width=0.67\columnwidth]{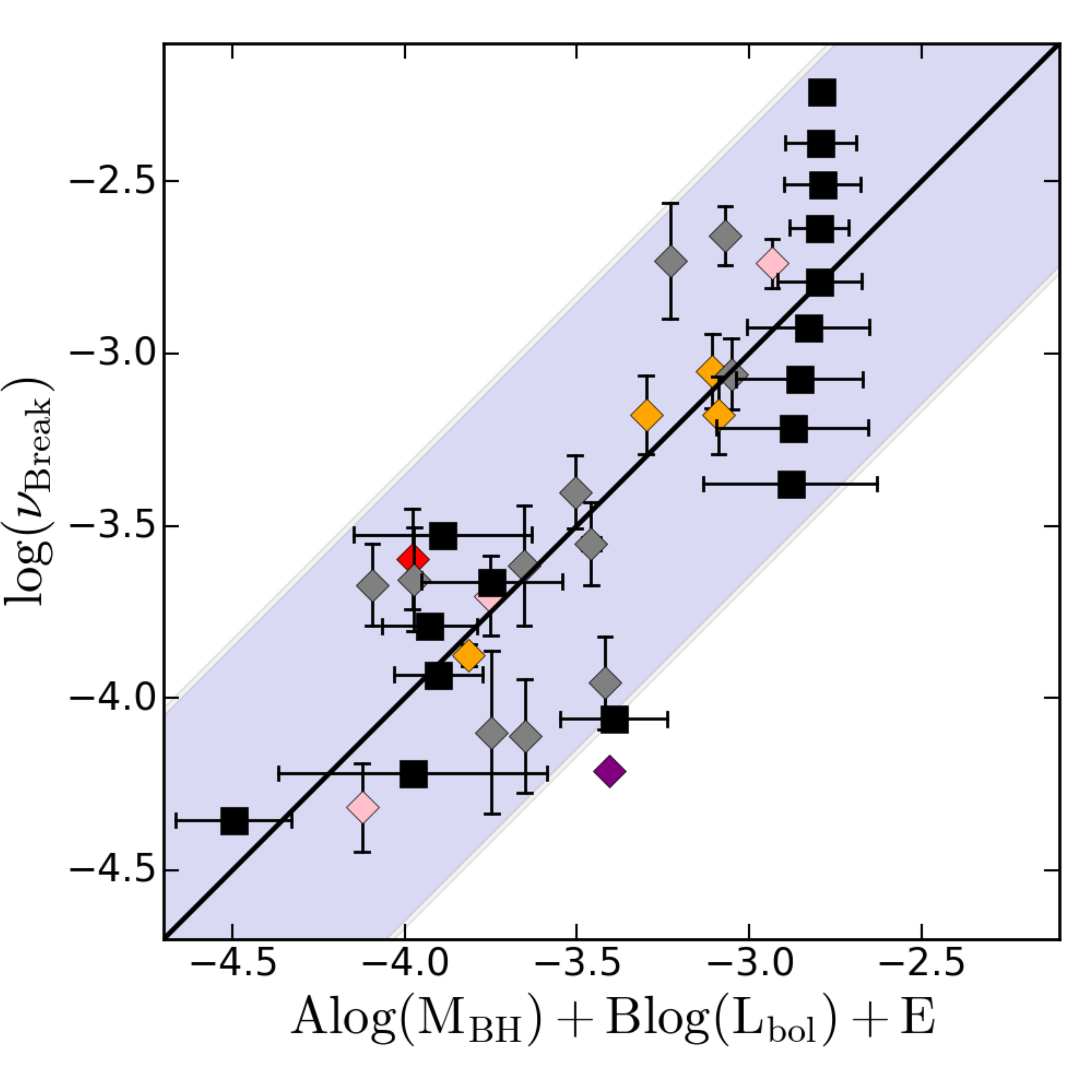}
\includegraphics[width=0.67\columnwidth]{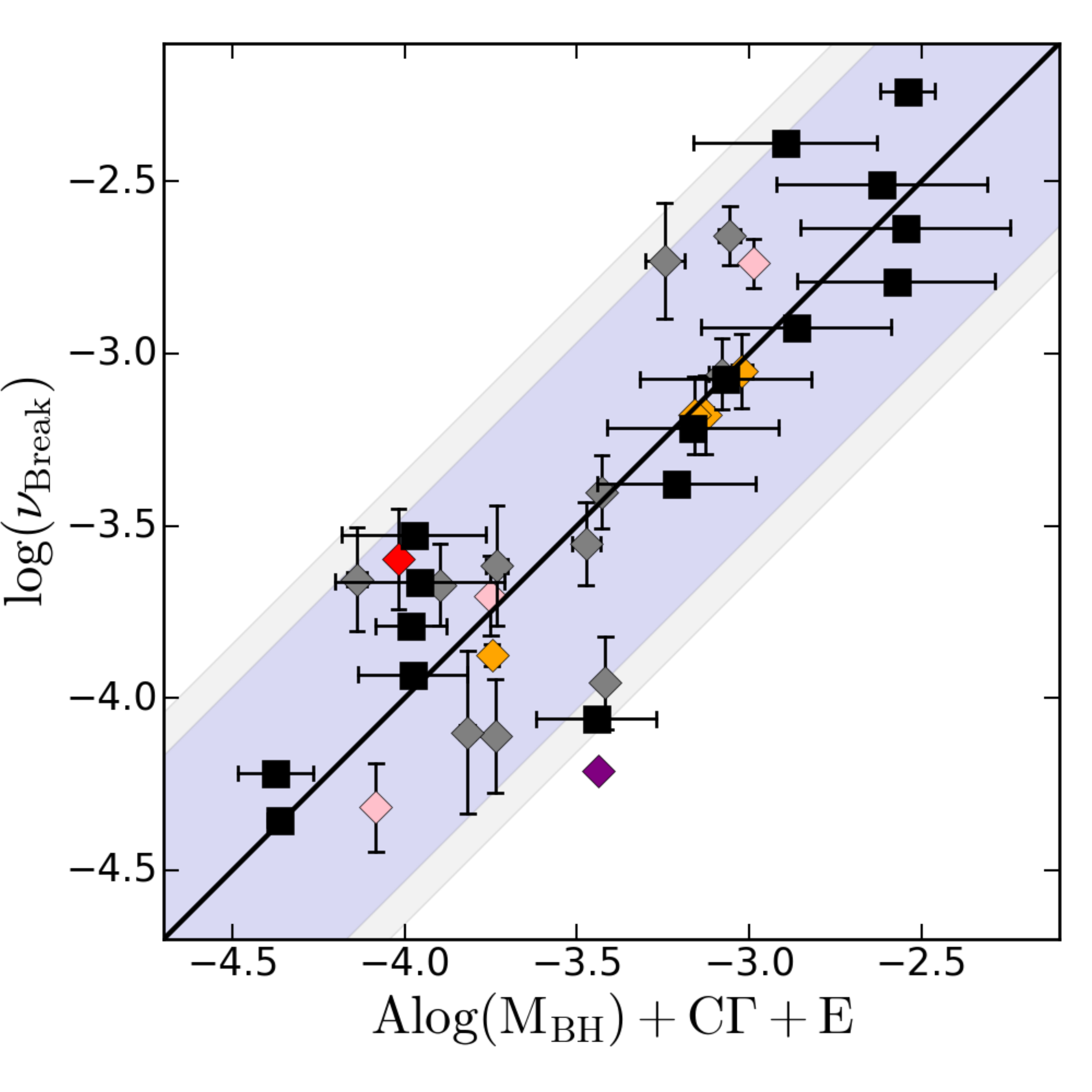}
\caption{Best-fit variability planes for the \emph{break-mass} relation of the form $\rm{log(\nu_{Break})=}$Alog($\rm{M_{BH}}$) + E (left), the \emph{break-mass-luminosity} relation of the form $\rm{log(\nu_{Break})=}$Alog($\rm{M_{BH}}$) + B log($\rm{L_{bol}}$) + E (middle); and the \emph{break-mass-index} relation of the form $\rm{log(\nu_{Break})=}$Alog($\rm{M_{BH}}$) + C$\rm{\Gamma}$ + E (right). Diamonds show the average value per object and black squares show the average value per interval of break frequencies. Diamonds are shown in different colors attending to the BH mass estimate method: reverberation mapping (gray), 5100$\rm{\AA}$ luminosity (orange), m-sigma relation (red), radio fundamental plane (pink), and water masers (purple). Gray and blue shadowed areas display two standard deviations when averaging per object or per break frequency intervals, respectively. Note that each plot is made using the maximum number of break frequencies for each of them (i.e. 2021 segments for the right panel and 2009 segments for the middle and left panel). }
\label{fig:plane}
\end{center}
\end{figure*}

Table~\ref{tab:plane} shows the best-fit results of each plane. Each plane is computed twice: (1) averaging all the breaks for a single object and (2) averaging in intervals of break frequencies. The latter allows to avoid biases on the detected break frequencies in our sample due to the limitation of the frequency range. We perform 100 Monte-Carlo simulations to take into account individual errors for each segment. Furthermore, we also run 100 Monte-Carlo simulations to take into account the error on the final parameters due to the standard deviation over the mean (per object or per frequency interval). This takes into account the dispersion of the measurements per interval. We use the average error instead of the standard deviation for the break-mass-nhvar relation. Table~\ref{tab:plane} includes the errors of the final parameters as the standard deviation of the 10,000 Monte-Carlo simulations run for each variability plane. 

We have recorded 2063 break frequencies among our sample of 22 AGN. All except ESO\,113-G010 have good estimates on the BH mass. This implies that we have 2021 reported PSD break frequencies with associated BH mass. Among those, the X-ray luminosity or the spectral index was estimated for 2009 segments and the absorption along the line of sight for 739 segments. Note that 739-segments subsample is included in the 2009-segments subsample with available X-ray luminosity and spectral index. We use these three sub-samples composed of 2021, 2009, and 739 segments, respectively, to determine if a more complex variability plane is indeed statistically required by the data using f-test. The break-mass relation (Eq. \ref{eq:bh}) was evaluated three times (one per subsample of segments), the break-mass-luminosity (Eq. \ref{eq:bhlbol}) and break-mass-index (Eq. \ref{eq:bhgamma}) relations were evaluated twice for the 2009- and 739-segments subsamples (since bolometric luminosity and spectral index are required parameters), and the break-mass-nh and break-mass-nhvar relations (Eqs. \ref{eq:bhnh} and \ref{eq:bhnhvar}) were evaluated for the 739-segments subsample for which the absorption is determined. This procedure enables to compare complex models with simpler ones. The f-test probability is included in Col.~11 of Table~\ref{tab:plane}. Ticks identify those best fits that are significantly better (i.e. f-test probability below $\rm{10^{-4}}$) than the prior (less complex) model for the same subsample.

Fig.~\ref{fig:plane} shows the best fit for each model using the maximum number of plausible segments for the break-mass, break-mass-luminosity, and break-mass-index relations. The slope \emph{A} associated with the BH mass is very well constrain for most of the models proposed with values well above the computed error (see Table~\ref{tab:plane}). The relation between the break frequency and the BH mass is then reinforced by our analysis. However, any relation between the break frequency and the bolometric luminosity is ruled out by our analysis (although it can be recovered with careful selection of the segments, see below). Only using the 2009-segments subsample and averaging in break-frequency intervals the break-mass-luminosity relations is preferred against the break-mass relation, although the slope \emph{B} associated with the bolometric luminosity is consistent with zero within two standard deviations. Slightly better results are obtained for the break-mass-index relation where using the 2009-segment subsample and averaging in break-frequency intervals, giving a slope \emph{C} associated with the spectral index well constrained. Neither of these two models are required by the data when we use the 739-segment subsample. 

\begin{figure*}
\begin{center}
\includegraphics[width=1.0\columnwidth]{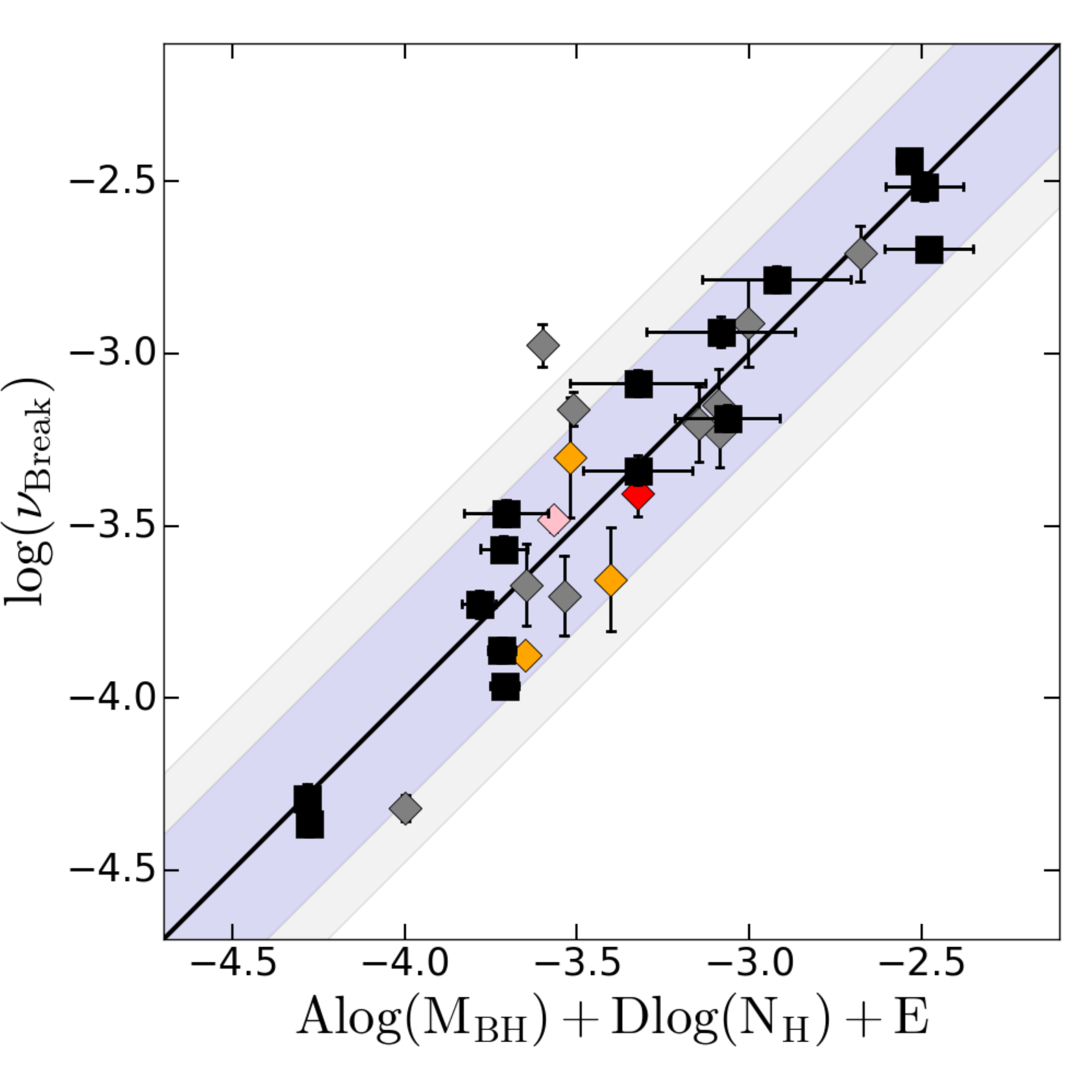}
\includegraphics[width=1.0\columnwidth]{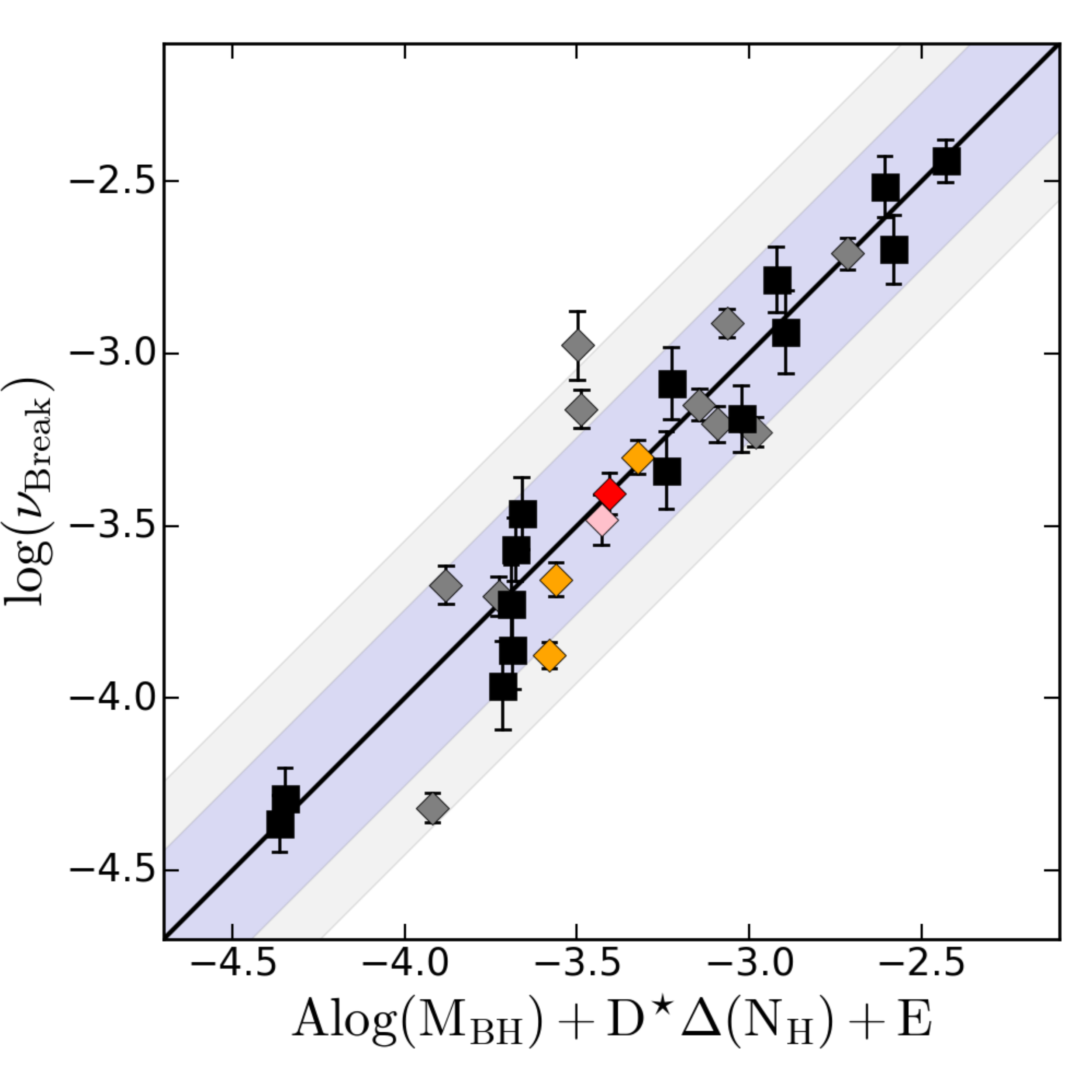}
\caption{Best-fit variability planes for the  \emph{break-mass-nh} relation of the form $\rm{log(\nu_{Break})=}$Alog($\rm{M_{BH}}$) + D log($\rm{N_{H}}$) + E (left) and the \emph{break-mass-nhvar} relation of the form $\rm{log(\nu_{Break})=}$Alog($\rm{M_{BH}}$) + $\rm{D^{\star} \Delta(N_{H})}$ + E (right). Symbols are described in Fig.~\ref{fig:plane}. Note that these plots are built using the maximum number of break frequencies with reported absorptions (739 segments). }
\label{fig:plane_nh}
\end{center}
\end{figure*}

The best variability plane is obtained when we involve the hydrogen column density in to the relation. Fig.~\ref{fig:plane_nh} shows the best fit for each model using the 739-segments subsample for the break-mass-nh (Eq.~\ref{eq:bhnh}) and break-mass-nhvar (Eq.~\ref{eq:bhnhvar}) relations. Indeed, the result is solid both averaging per object or per frequency break intervals and the slopes \emph{D} and \emph{D$^{\star}$} (associated with the absorption and absorption variations, respectively) are well constrained according to the associated errors. We cannot statically prefer one of the two scenarios. Due to the underlying bias of averaging by object (since we do not have all the frequency break per object) we suggest the reader to use the break-mass-nh or break-mass-nhvar relation found by averaging in intervals of frequencies, i.e: 
\begin{equation}
\footnotesize
log(\nu_{Break})= -0.59~log(M_{BH}) +0.10~log(N_{H})  - 1.5 \label{eq:finalbhnh} 
\end{equation}
\begin{equation}
\footnotesize
log(\nu_{Break})= -0.55~log(M_{BH}) +0.56~\Delta(N_{H})  + 0.2 \label{eq:finalbhnhvar}
\end{equation}

\begin{figure}
\begin{center}
\includegraphics[width=1.0\columnwidth]{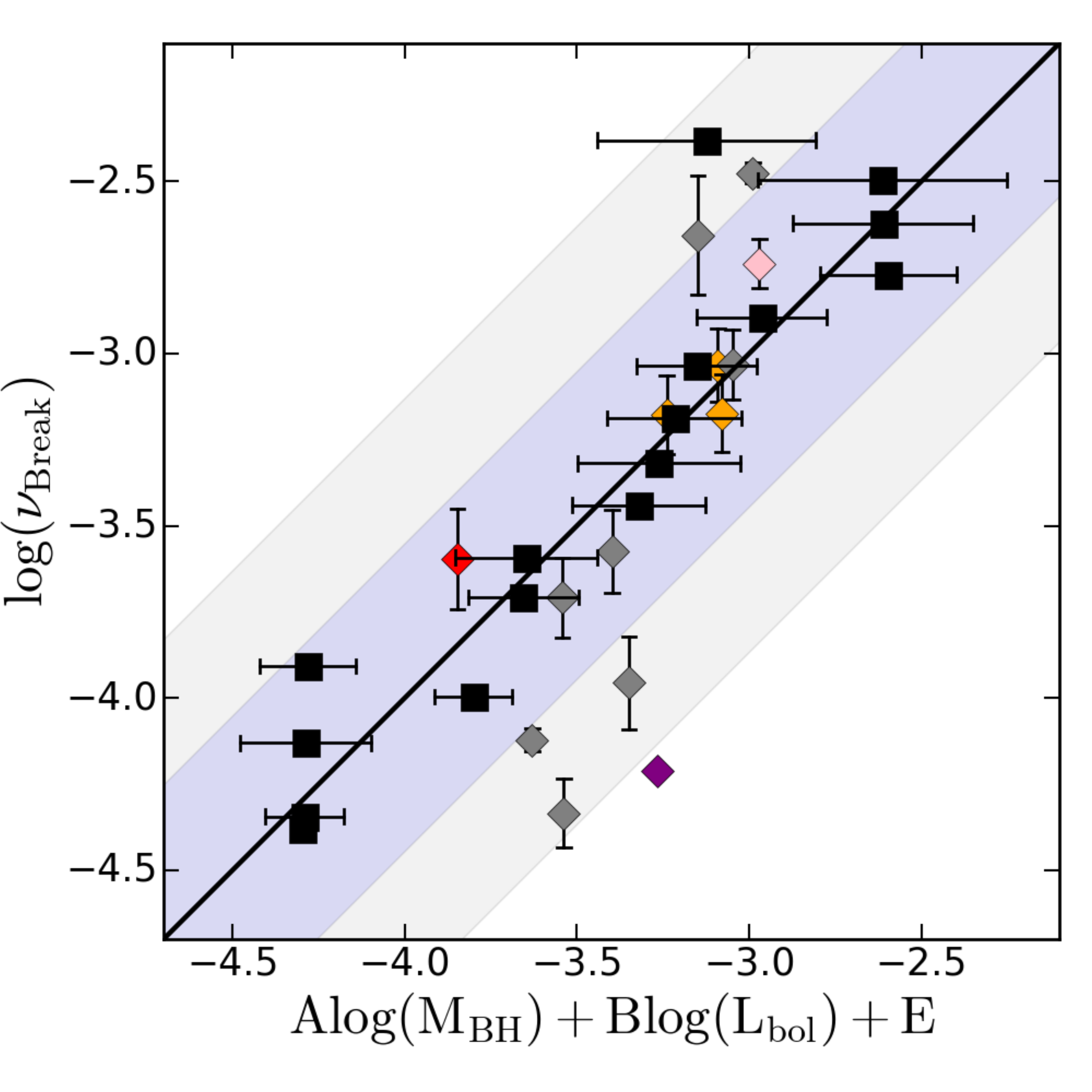}
\caption{Best-fit for the break-mass-luminosity relation ($\rm{log(\nu_{Break})=}$Alog($\rm{M_{BH}}$) + B log($\rm{L_{bol}}$) + E) for non-absorbed segments (see text). Symbols are described in Fig.~\ref{fig:plane}.}
\label{fig:plane_noObscured}
\end{center}
\end{figure}

We also tried more complex models including at the same time the bolometric luminosity and absorption, spectral index and absorption, and absorption and absorption variations. However, none of them were statistically needed. 

Our result implies that absorption along the line of sight plays a major role in the determination of the break frequency. This is related to clouds passing by our line of sight and close enough to the observer to produce noticeable absorption variations. Nevertheless, could we recover any relationship between the PSD break frequency and the bolometric luminosity when excluding such events? We tackle this issue in two ways. First, we performed again our analysis using only segments where absorption was not determined, under the assumption that those intervals correspond to very low absorption or not absorbed segments at all. This new re-analysis contained 1270 segments with estimates on the BH mass and the bolometric luminosity. Indeed, this new analysis shows a significant improvement when including the bolometric luminosity or the spectral index into the variability plane (both of them with f-test probability well below $\rm{10^{-4}}$, see Table~\ref{tab:plane}). We plot in Fig.~\ref{fig:plane_noObscured} the best fit break-mass-luminosity relation obtained for these non-absorbed segments. The improvement on the relation compared to Fig.~\ref{fig:plane} (middle panel) can readily be spotted. 

The second test takes the advantage that our sample seems to have a subset of eight objects that do not show absorption or their absorption do not seem to be variable (see Section \ref{sec:perobject}). We re-analyzed our models using this sample of eight sources (see Table~\ref{tab:plane}). We found that both bolometric luminosity and the spectral index could play a role on the variability plane. The plane is significantly better than the break-mass relation, but they produce less accurate results compared to the break-mass-absorption relation to the entire sample and the break-mass-luminosity relation obtained using segments without absorption. Therefore, if no absorber (in particular no variable absorber) is found, our results are consistent with the following variability plane: 
\begin{equation}
\footnotesize
log(\nu_{Break})= -1.39~log(M_{BH}) +0.82~log(L_{bol})  - 30.0 \label{eq:finalbhlbol}
\end{equation}

Just as a sanity check we also used AGN with variable absorption (i.e. those that show variations on the $\rm{N_H}$ parameter, see Fig.~\ref{fig:variability} and Section~\ref{sec:perobject}), to reproduce the five variability planes (Eqs. 4-8), recovering fully consistent results. Therefore, we found no indications of the need of the bolometric luminosity or the spectral index to explain variations in the PSD break frequency for AGN with detected (variable) absorber.  

Finally, we highlighted with different colors in Figs. \ref{fig:plane}-\ref{fig:plane_noObscured} objects with BH masses estimated with different methods. We do not see any particular trend.

\section{Discussion}\label{sec:discussion}

\subsection{Break-mass-nh relation}\label{sec:discussion_nh}

Our new variability plane shows that the break frequency scales with the BH mass and the (variability of the) obscuration along the line of sight (Eqs.~\ref{eq:finalbhnh} and  \ref{eq:finalbhnhvar}). Note that we do not find evidence that both obscuration and luminosity could play a role on the determination of the break frequency at the same time. On the contrary, only 739 obscured segments are able to destroy the break-mass-luminosity relation that can be recovered using the other 1270 unobscured segments (see Table\,\ref{tab:plane}). Thus, as long as there is enough obscuration, it seems to be the dominant process determining the break frequency.

Absorption is variable in timescales of hours for many of our objects (see Fig.~\ref{fig:variability} and Section \ref{sec:perobject}) so it must be then related to clouds close the accretion disk. The observed X-ray variability in type-1 AGN may be caused by variations in the covering fraction with a large portion of the flux obscured due to these clouds \citep{Done07B,Miller07,Miller09}. Eclipses by clouds are rather common in type-1 AGN  \citep[e.g. MRK\,766, NGC\,1365, or NGC\,4593][]{Miller07,Risaliti11,McHardy17}. 

We present here a toy model for clouds close to the accretion disk associated with a clumpy wind. Firstly lets assume that these clouds follow an exponential distribution of hydrogen column densities of the form $\rm{N_{H}(r) = N_{H,in}(r/r_{in})^{-\alpha}}$, where $\rm{N_{H,in}}$ is the $\rm{N_{H}}$ for individual clouds at the inner radius $\rm{r_{in}}$ of the cloud distribution. The radius of the innermost stable circular orbit $\rm{R_{ISCO}}$ depends on the BH spin, being at $\rm{R_{ISCO} = R_{g}}$ for Kerr BHs and $\rm{R_{ISCO} = 6 R_{g}}$ for Schwarzchild BHs, where the gravitational radius is $\rm{R_{g}=GM_{BH}/c^{2}}$. We assume that the cloud inner radius is also at a number $\rm{N_{g}}$ of gravitational radius $\rm{R_{g}}$ of the central SMBH:  $\rm{r_{in}=N_{g}R_{g} = N_{g}GM_{BH}/c^{2}}$. 

If we assume these clouds are governed by the SMBH gravitational potencial, i.e. they are moving in Keplerian orbits \citep[broadly assumed by the community for BLR clouds to estimate BH masses, e.g.][]{Peterson04}, the circular velocity of each cloud is $\rm{v_{circ}=(GM_{BH}/r)^{1/2}}$. The characteristic frequency associated to this orbit is $\rm{\nu_{orbit}=v_{circ}/r=(GM_{BH}/r^{3})^{1/2}}$. Using the radial profile distribution of the hydrogen column density and the definition of $\rm{r_{in}}$ (above) we can re-write it in terms of $\rm{N_{H}(r)}$:
\begin{equation}
\nu_{orbit}(r)= \bigg(\frac{c}{N_g}\bigg)^{3/2}\bigg(\frac{1}{GM_{BH}}\bigg)\bigg(\frac{N_{H}(r)}{N_{H,in}}\bigg)^{3/2\alpha} \label{eq:nuorbit}
\end{equation} 

Typical inner (outer) clouds radii associated to the BLR of $\rm{\sim}$10$\rm{R_g}$ ($\rm{\sim}$1,000$\rm{R_g}$) produce orbits with associated frequencies of $\rm{10^{-1}}$Hz ($\rm{10^{-6}Hz}$) and $\rm{10^{-4}}$Hz ($\rm{10^{-9}Hz}$), for BH masses of $10^5M_{\odot}$ and $10^8M_{\odot}$, respectively. These frequencies overlap with the frequency range presented in this analysis. This formalism gives a simple explanation for the break-mass-nh relation found, even showing positive slope associated to $\rm{N_H}$ ($\rm{\alpha>0}$) and negative slope associated to $\rm{M_{BH}}$. Tentatively, the slope associated to the $\rm{N_H}$ ($\rm{D\sim 0.1}$) imposes a steep profile for $\rm{N_{H}(r)}$ ($\rm{D=3/2\alpha}$).

The break-mass-nhvar relation is also well explained if the higher absorption seen at inner radii is due to a larger number of clouds. Inner clouds move faster than outer clouds. Thus, a higher number of clouds toward the inner radii (associated with high frequencies) naturally will show large $\rm{N_{H}}$ dispersion.

However, note that the radial profile presented here is oversimplified. Firstly, changes to the cloud number, cloud sizes, or obscuration per cloud along the radius all yield to a different $\rm{N_{H}(r)}$ column density profile. Thus, a steep $\rm{\alpha}$ could be obtained by a steep radial distribution of either cloud sizes, individual cloud obscuration, or number of clouds. Moreover, the $\rm{N_{H}(r)}$ radial profile could be more complex than a single power-law. For instance a broken power-law distribution of clouds could explain a bending power-law model for the PSD shape \citep[see ][]{Zhang17}. Finally, in order to account for the total number of frequencies in a particular time, we need to take into account that the number of clouds at a given radius should be distributed randomly along Kepler orbits (i.e. depending on polar angle $\rm{\theta}$) and then this initial stochastic distribution of cloud will evolve in time, following $\rm{\theta(t)=\Omega_{circ} t}$, where $\rm{\Omega_{circ}}$ is the angular velocity ($\rm{\Omega_{circ}=v_{circ}/r}$). These refinements prevent us for a proper comparison of the slopes found in this paper and those derived from Eq. \ref{eq:nuorbit}. Indeed, \citet{Zhang17} recently studied a few PSDs from simulated light-curves as a sum of eclipsing events. They show that absorption variations could reproduce the bending PSD found in AGN and it also explains the break-mass relation (see their Fig.~10). They naturally reproduce a frequency break if there is a bend in the spatial distribution of clouds, if no eclipsing clouds exist within a certain radius, or if the size of the eclipsing clouds compared to the X-ray source scales with the obscuration. Furthermore, they found that the break frequencies might depend on the BH mass, the size of the X-ray emitting region, the effective radius of the eclipsing clouds, and the distribution of clouds along the radius. Unfortunately, \citet{Zhang17} did not explore a relationship between break frequency and absorption although they reported changes on the break frequency associated with the parameters of the cloud distribution. We have found that the break-mass-relation is dominant over the break-mass-luminosity relation. An open question question is why this happens. In order to tackle this issue, we need to produce a model including the accretion process together with eclipsing events. 

A caveat to keep in mind is that these results are based in a certain spectral modeling used to infer the $\rm{N_H}$. Seven of these AGN have been claimed to have a broad red-shifted iron line $\rm{FeK\alpha}$ associated to the reflection of the X-ray photons in the disk \citep{Fabian03,Reynolds14,Marinucci14,Kammoun17} although absorption-dominated model using a clumpy absorber could also explain these spectral features \citep{Miller09}. If relativistic effects are occurring in general, the spectral modeling performed here is not longer appropriate. The reflection model is quite complex, requiring high energy bands and very high S/N ratio, which is not viable for our short segments of time. The extension of the red-shifted wing of the $\rm{FeK\alpha}$ line depends on the innermost stable orbit \citep[$\rm{R_{ISCO}}$,][]{Reynolds14}. Thus, our changes on the absorption along the line of sight could mimic changes on the $\rm{R_{ISCO}}$. A different $\rm{R_{ISCO}}$ has been claimed to explain some of the dispersion seen in the break-mass relation \citep{Vaughan03}. However, $\rm{R_{ISCO}}$ is a function of the spin \citep{Bardeen72} which is not expected to vary for a single object \citep{Vaughan03}. Thus we rule out this scenario. 

\subsection{Break-mass-luminosity relation}\label{sec:discussion_lum}

We explore here how the break-mass-luminosity relation compares with previously found relations and why other authors found it as a primary variability plane while here we only obtained it after filtering obscured segments (break-mass-nh/nhvar relationships are preferred otherwise). Indeed note that the number of obscured segments (739) is significantly lower than the number of unobscured segments (1270). However, the break-mass-nh relation is preferred when we use all of them together (2009, see Table\,\ref{tab:plane}). Furthermore, we did not get a significantly better result when a combination of BH mass, bolometric luminosity, and obscuration was tested (see Section \ref{sec:results}). Thus, obscuration is dominant.

Under the break-mass-luminosity relation, these break frequencies are associated with the inner edge of the accretion disk, moving closer to the BH as systems change from the low to the high state mainly due to changes in the accretion rate \citep{Lyubarskii97,McHardy04} or changes in the location of the last stable orbit due to a different spin \citep{Vaughan03}.

The break-mass-luminosity relation is mostly expressed in the literature in terms of timescale (i.e. 1/$\rm{\nu_{Break}}$) in units of days, the BH mass in units of $\rm{10^{6}M_{\odot}}$, and the bolometric luminosity in units of $\rm{10^{44}erg/s}$. Our break-mass-luminosity relation in this form can be written as: 
\begin{equation}
\footnotesize
log(t_{Break})= 1.39~log(M_{BH}) -0.82~log(L_{bol})  - 2.7 \label{eq:finalbhlbol_t}
\end{equation}

We compare this relationship with the following variability planes:
\begin{equation}
\footnotesize
log(t_{Break})= 2.2~log(M_{BH}) -0.9~log(L_{bol})  - 2.4 \label{eq:mchardy} 
\end{equation}
\begin{equation}
\footnotesize
log(t_{Break})= 1.1~log(M_{BH})  - 1.7 \label{eq:ogm12_1} 
\end{equation}
\begin{equation}
\footnotesize
log(t_{Break})= 1.3~log(M_{BH}) -0.2~log(L_{bol})  - 1.9 \label{eq:ogm12_2}
\end{equation}

\noindent where Eq. \ref{eq:mchardy} is the first attempt to quantify the break-mass-luminosity relation by \citet{McHardy06} and Eqs. \ref{eq:ogm12_1} and \ref {eq:ogm12_2} are the break-mass and break-mass-luminosity relation reported by \citet{Gonzalez-Martin12}. Note that the errors on the parameters are of the order of $\rm{\sim 0.3}$.

\citet{McHardy06} show a slope associated to the BH masses significantly steeper (A$\rm{\sim}$2.2) than our measurement (A$\rm{\sim}$1.4). However, our value is fully consistent with the slope found by \citet{Gonzalez-Martin12} for both the break-mass and break-mass-luminosity relation. The slope associated to the bolometric luminosity reported here is fully consistent with \citet{McHardy06}. 

\citet{Gonzalez-Martin12} were not able to find a relation between the break frequency and the bolometric luminosity. This is easily explained because \citet{Gonzalez-Martin12} used almost the same observation analyzed here. Thus, most of these objects show breaks that are related to the absorption (variations) rather than to changes in the bolometric luminosity. In favor, the scenario of eclipsing clouds proposed by \citet{Zhang17} does not produce any strong relation with the Eddington ratio. 

However, how \citet{McHardy06} found this relation if the frequencies of eclipsing clouds are the dominant behavior in our analysis? Whatever causes the break-mass-nh or the break-mass-nhvar relations, it did not affect most of the break frequencies in their analysis. We believe that they used break frequencies less affected by eclipsing clouds. A clean path is obtained stochastically when no clouds are intercepted along the line of sight. Isolating these events naturally recovers the underlaying break-mass-luminosity relation, associated with the accretion process \citep[reported in Eq. \ref{eq:finalbhlbol} and previously found by][]{McHardy06}. We guess that \citet{McHardy06} found the break-mass-luminosity relation because they relied mostly in high frequency breaks. Among their primary sample of 10 AGN, they used high break frequencies (above our frequency range) in six objects. High break frequencies are less affected by orbital frequencies of the clouds because they would be associated with external clouds that are less frequent in the system (due to the decrease of the number of clouds along the radius). 

Therefore, break frequencies seem to be associated to accretion processes or eclipsing clouds. At the frequency range studied in our paper, break frequencies are associated to accretion processes only when those associated to eclipsing clouds are excluded. We actually do not find evidence that the break frequencies could be associated to both, bolometric luminosity and obscuration at the same time. However, a proper PSD model of an accretion disk surrounded by a cloud distribution is needed to understand which conditions are required for one of these two mechanisms dominate over the other one. This will be the subject of a future investigation.

\subsection{BH masses through the variability plane}\label{sec:discussion_bhm}

Leaving the physical interpretation of the variability plane found in this work aside, this relation can be used to infer BH masses when these are not available using more direct methods. Among the objects in our sample, ESO\,113-G010 is a good example of no BH mass estimate reported before. 

ESO\,113-G010 do not seem to show any variations on the absorber, which seems to be low, considering that we detect the absorption for only one of the segments. Thus, the break-mass-luminosity relation seems to apply in this case (Eq. \ref{eq:finalbhnh}). We computed the BH mass estimate using the 41 break frequencies found for this object\footnote{We excluded one of the detected PSD frequency breaks to avoid segments where absorption along the line of sight was detected.}. For each of these breaks, we used the Monte-Carlo method to include the errors on the parameters of the break-mass-luminosity relation and the errors on the luminosity and break frequencies. We found a BH mass estimate of $\rm{log(M_{BH})=6.8\pm0.2}$.

Optical reverberation mapping has established a scaling relationship between the BLR radius and AGN luminosity \citep{Kaspi00,Bentz09} which allows single-epoch BH mass estimates when combined with the measurement of the broad line width. This method was used by \citet{Cackett13} to estimate a BH mass of $\rm{log(M_{BH})}$=6.8 for ESO\,113-G010, fully consistent with our result.

Furthermore, \citet{Porquet07} estimate the BH mass in ESO\,113-G010 using the break-mass-luminosity relation reported by \citet{McHardy06} finding $\rm{log(M_{BH})}$=[6.6-7.0], which is also consistent with our result. This reinforces that the variability plane recovered for non-absorbed segments is fully consistent with that previously found by \citet{McHardy06}. 

We also estimate (using the same methodology) the BH mass using the break-mass-nh plane reported here for the first time. For that purpose, we use the single segment with both frequency break and $\rm{N_H}$ measured. However, our estimate on the BH mass using the break-mass-absorption relation is quite below the previous estimates ($\rm{log(M_{BH})=6.0\pm0.1}$). Thus, we warn the reader to use the break-mass-nh relation only when the absorption is clearly showing variations in short time scales. This guarantees that the clouds are located close to the nucleus so they are related to the observed PSD break frequency. An advantage of this relation is that, as long as we can guarantee that the clouds are close enough to the central BH, this relation could be used to infer BH masses, irrespective of the AGN type since X-rays could penetrate large amount of obscuration. Thus, this method could be used for type-2 AGN for which reverberation mapping techniques are not suitable, together with other emerging techniques \citep[e.g. the profile of the FeK$\alpha$ line,][and references therein]{Minezaki15}. At the same time we would like to underscore that the break-mass-luminosity relation must be used carefully. As shown in this paper, this relation remains valid only when a clean view of the inner corona can be guaranteed. The structural information might be washed out by cloud motions, otherwise.

\section{Summary}\label{sec:summary}

We analyze the light-curves of a sample of 22 luminous AGN (mostly Sy1 and NLSy1) observed with \emph{XMM}-Newton with previously reported break frequencies in their integrated PSD. The main difference with previous analysis is that we divide the light-curves in short segments to study the behavior of the PSD break frequency in short timescales. The main results are:
\begin{itemize}
\item The PSD break frequency is not a unique quantity defined by each source. It varies in 19 out of the 22 AGN (i.e. $\rm{\sim}$ 90\%).
\item Using over two thousand frequency breaks, we found a new variability plane, which links the break frequencies to the BH masses and the absorption along the line of sight (break-mass-nh relation). Alternatively, we also found a good fit of the reported break frequencies using a combination of the BH masses and the absorption variations (break-mass-nhvar relation).
\item The previously found variability plane that links break frequency with BH mass and bolometric luminosity (break-mass-luminosity relation) is only recovered when absorption is not found along the line of sight. This suggests that, among our sample, there are frequency breaks associated to the absorption while others are probably associated to accretion processes. Frequency breaks associated to absorption processes are dominant in the frequency range studied in this paper.
\end{itemize}

Our new variability plane might be explained if the frequencies are related to the orbits of individual clouds, although a comprehensive theoretical model, including realistic cloud distribution and also combining with the PSD produced by both clouds and the accretion disk, is still lacking. We leave this to a further investigation. This scaling relation might have the power to set constraints on the distribution of the clouds since any proposed distribution should recover a slope of $D\sim 0.1$ associated to the $\rm{N_{H}}$. Proper simulations trying to reproduce the light-curve and, at the same time, the reported scaling relations might put strong constraints on the size, distribution, and distance of the absorber. Although emission line reverberation mapping techniques also have this power \citep{Peterson04}, it is important to stress that the study of the $\rm{N_{H}}$ variability throughout X-ray data requires significantly less telescope time. 

\acknowledgments

I thank to the anonymous referee for his/her useful comments. This work is mostly funded by UNAM PAPIIT projects IA 100516 and IA 103118. This paper is the result of fruitful previous researches with S. Vaughan and I. Papadakis. I also thank to A. Negrete and E. Benitez for useful information about black hole mass estimates and D. Dultzin and J. Masegosa for long-standing discussion on the AGN structure. Finally, I would like to thank also to the AGN group at IRyA (M. Martinez-Paredes, A. Pasetto, D. Esparza-Arredondo, N. Osorio-Clavijo, and C. Victoria-Ceballos) for our group meeting discussions that inspired this research. 

\appendix

\section{Selection of black-hole mass estimates}\label{sec:BHmass}

BH mass estimates are key in our analysis. Unfortunately, different methods give inconsistent results for the same object. The size of the discrepancy is difficult to spot since there are no galaxies for which BH masses have been independently confirmed using more than one technique \citep{Merritt01}. We compile BH masses with different methods in Table \ref{tab:BHmass}. Discrepancies up to more than an order of magnitude are seen in some of our objects. Indeed, most of the objects show discrepancies. Although we have done our best to avoid systematics on the BH mass selection (see below), we warn the reader about the unavoidable uncertainties associated to this selection.

One of the best methods are spatially resolved kinematics. However, only a few dozen of objects have SMBH mass estimates based on spatially resolved kinematics due to complex behavior in the central few tens of parsecs of the galaxies \citep[see][for a review]{Kormendy01}. Furthermore, masses derived from ground-based stellar kinematics are roughly an order of magnitude above than those inferred from other technique \citep{Merritt01}. Reverberation mapping use time delays between brightness variations in the continuum and in the broad emission lines, interpreted as the light travel time between the SMBH and the line-emitting region farther out \citep{Blandford82,Netzer97}. Reverberation mapping is the best way to estimate SMBH masses for AGN. We used masses derived with this method in 12 objects. Note that we excluded some reverberation mapping measurements due to the lack of measured errors (as far as we could track). Some objects show several BH mass estimates using reverberation mapping but they do not agree among them. Thus, we choose the author \citet[][]{Zu11} in 9 out of these 12 objects to avoid systematics errors due to the methodology applied. 

\citet{Kaspi00} obtained an empirical relation between the BLR size of the AGN from the reverberation mapping sample and the optical continuum luminosity at 5100$\rm{\AA}$ which can be used to estimate BH masses. We used this estimated in other five sources when available. 

The m-sigma relation is an empirical correlation between the stellar velocity dispersion of a galaxy bulge and the SMBH mass \citep{Ferrarese00}. We used this relation for another two sources. We used the radio fundamental plane \citep{Laor00,Gliozzi10} and masers \citep{Graham08} for the last two sources where no other measurement were available. As mention in the main body of the text, ESO\,113-G010 has no BH mass measurements reported. 

\begin{table*}[t!]
\scriptsize
\begin{center}
\caption{Compilation of BH masses.}
\begin{tabular}{lccc|lccc}
\toprule
Object        & $\rm{log(M_{BH})}$  	& Method & Reference		& Object         		& $\rm{log(M_{BH})}$ 	& Method & Reference		  \\
	(1)		  &			(2)			&	 (3)    &			(4)		& 		(1)	     		&		(2)			   	&	   (3)    &	(4)		  \\ \hline \hline
MRK\,335 	  & {\bf 7.23$\pm$0.04} 	& R  & \citet{Zu11}			& MRK\,766 	  	& {\bf 6.2$\pm$0.3}   	& R  & \citet{Bentz09}	  \\
	   	 	  & 7.4$\pm$0.1  		& R	 & \citet{Grier12} 		& 			  		& 6.57 			   		& S  & \citet{Botte05} 	  \\  
	   	 	  & 6.93$\pm$0.1 		& R	 & \citet{Du15} 			& 			     		& 6.5$\pm$0.3 		   	& S  & \citet{Uttley05}	  \\ 
			  & 7.34					& X  & \citet{Arevalo08}		& NGC\,4395      	& {\bf 5.5$\pm$0.1}   	& R  & \citet{Peterson05}  \\  
	     	  	  & 7.15$\pm$0.12 		& R  & \citet{Peterson05} 	& 			     		& 4.70 			   		& X  & \citet{Vaughan05B} \\ 
ESO\,113-G010 &		\dots		&\dots	 &		\dots		&			     		& 5.60 			   		& K  & \citet{denBrok15}   \\   
Fairall\,9	  & {\bf 8.3$\pm$0.1}   	& R  & \citet{Zu11}			&  			     	       & 5.23 			   		& R  & \citet{LaFranca15} \\  
			  & 8.41		 			& R  & \citet{Peterson05} 	& MCG\,-06-30-15   & {\bf 6.3$\pm$0.4}   	& L  & \citet{Zhou10}	    \\
			  & 8.09$\pm$0.1 		& R  & \citet{Du15} 			& 			     		& 6.71 			   		& X  & \citet{McHardy05}   \\
PKS\,0558-504 & {\bf 8.48$\pm$0.05} & RP & \citet{Gliozzi10}	& IC\,4329A 	 	& {\bf 8.3$\pm$0.5}   	& S  & \citet{Markowitz09} \\ 
1H\,0707-495  & {\bf 6.3$\pm$0.5} 	& L	 & \citet{Bian03} 		& 			     		& 6.99 			   		& R	  & \citet{Peterson05}  \\
			  & 6.37  				& R	 & \citet{Zhou05}		& Circinus	     		& {\bf 6.04$\pm$0.08} 	& M	  & \citet{Graham08}   \\  
ESO\,434-G40  & {\bf 7.6$\pm$0.2} 	& S	 & \citet{Peng06}		&			     		& 6.23$\pm$0.1		   	& M  & \citet{Greenhill03} \\ 
NGC\,3227 	  & {\bf 6.8$\pm$0.1}   	& R  & \citet{Zu11}			& NGC\,5506 	 	& {\bf 8.1$\pm$0.2 }		   	& R	  & \citet{Du15}		  \\
			  & 6.88 				& R  & \citet{Denney09} 	& 			     		& 7.45 			   		& R	  & \citet{Wang07}	  \\   
			  & 7.09$\pm$0.1 		& R  & \citet{Du15}			& 					& 8.0$\pm$0.5   	& S  & \citet{Uttley05}	  \\ 
REJ\,1034+396 & {\bf 6.6$\pm$0.3} & L  & \citet{Zhou10} 		&				 	& 6.88 			   		& S	  & \citet{Peng06}	  \\ 
			  & 6.40 				& S	 & \citet{Bian09} 		& 			     		& 6.7$\pm$0.7 		   	& S	  & \citet{Bian09}	  \\   
NGC\,3516 	  & {\bf 7.40$\pm$0.05} 	& R	 & \citet{Zu11} 			& NGC\,5548 	 	& {\bf 7.72$\pm$0.02} 	& R	  & \citet{Zu11} 		  \\
			  & 7.50$\pm$0.05 		& R  & \citet{Denney09}		& 			     		& 7.64 			   		& R	  & \citet{Denney09}	  \\
			  & 7.82$\pm$0.05		& R  & \citet{Du15}			& NGC\,6860 	 	& {\bf 7.6$\pm$0.5}   	& L	  & \citet{Wang07}	  \\  
NGC\,3783 	  & {\bf 7.37$\pm$0.08} 	& R  & \citet{Zu11} 			& ARK\,564 	     	& {\bf 6.3$\pm$0.5}    	& L	  & \citet{Zhang06}	  \\
			  & 7.47 				& R	 & \citet{Peterson04}	&			  	 	& 6.0 				   	& R  & \citet{Shapovalova12} \\ 
			  & 7.45$\pm$0.1 		& R	 & \citet{Du15}			&			  	 	& 6.23 			   		& R	  & \citet{Denney09}	  \\ 
NGC\,4051 	  & {\bf 6.1$\pm$0.1} 	& R  & \citet{Zu11} 			&			  	 	& 7.00  			   		& X	  & \citet{Pounds01}	  \\ 
			  & 6.23 				& R	 & \citet{Denney09} 		& NGC\,7469 	 	& {\bf 6.96$\pm$0.05} 	& R   & \citet{Zu11}		  \\
			  & 5.42$\pm$0.25		& R  & \citet{Du15}			&			  	 	& 6.9$\pm$0.8		   	& R   & \citet{Du15}		  \\
NGC\,4151 	  & {\bf 7.55$\pm$0.05} 	& R  & \citet{Zu11} 			& 			     		&					   	&	   &					  \\		
			  & 7.12 				& R  & \citet{Peterson05} 	& 			     		&					   	&	   &					  \\
			  & 7.72$\pm$0.06  		& R	 & \citet{Du15}			& 			     		&					   	&	   &					  \\ 
\bottomrule
\end{tabular}
\end{center} 
\tablecomments{BH masses are taken from reverberation mapping techniques (R), m-sigma relation (S), continuum at 5100$\rm{\AA}$ (L), water masers (M), radio fundamental plane (RP), X-ray variability plane (X), and kinematics (K). BH masses used along this paper are highlighted with bold-face letters.}
\label{tab:BHmass}
\end{table*}

\end{document}